 \documentclass[11pt]{article}

\usepackage[english]{babel}
\usepackage{blindtext}
\usepackage{amsmath,amsfonts,amsfonts,amsthm}
\usepackage{cleveref}

\usepackage{graphicx}
\usepackage{amssymb}
\usepackage{bbm}
\usepackage{subfigure}

\usepackage[version=3]{mhchem}

\usepackage{lineno}
\usepackage[titletoc,title]{appendix}
\usepackage{algorithm,algpseudocode}
\usepackage{siunitx}
\usepackage{authblk}
\usepackage[square,sort,comma,numbers,compress]{natbib}

\theoremstyle{definition}
\newtheorem{definition}{Definition}[section]

\numberwithin{equation}{section}

\newcommand\chem[1]{\ce{#1}}

\title{Efficient time stepping for reactive turbulent simulations with stiff chemistry}

\author[]{Hao Wu\thanks{wuhao@stanford.edu}}
\author[]{Peter C. Ma\thanks{peterma@stanford.edu}}
\author[]{Matthias Ihme\thanks{mihme@stanford.edu}}
\affil[]{Department of Mechanical Engineering, \\Stanford University, Stanford, CA 94305, United States}

\begin{document}

\maketitle

\begin{abstract}
A combination of a steady-state preserving operator splitting method and a semi-implicit integration scheme is proposed for efficient time stepping in simulations of unsteady reacting flows, such as turbulent flames, using
detailed chemical kinetic mechanisms. The operator splitting is based on the \textit{Simpler} balanced splitting method, which is constructed with improved stability properties and reduced computational cost. The method is shown to be capable of stable and accurate prediction of ignition and extinction for reaction-diffusion systems near critical conditions. The ROK4E scheme is designed for semi-implicit integration of spatially independent chemically reacting systems. Being a Rosenbrock-Krylov method, ROK4E utilizes the low-rank approximation of the Jacobian to reduce the cost for integrating the system of ODEs that have relative few stiff components. The efficiency of the scheme is further improved via the careful choice of coefficients to require three right-hand-side evaluations over four stages. Combing these two methods, efficient calculation is achieved for large-scale parallel simulations of turbulent flames. 
\end{abstract}



\section{Introduction}
\label{S:introduction}

High fidelity simulations of reactive turbulent flows using finite-rate chemistry is a promising tool for furthering the understanding of physical processes such as ignition, extinction, flame stabilization, pollutant formations, etc.~\cite{WU_SEE_WANG_IHME_CF2015,lignell2007effect,jaravel2017large,yang2017sensitivity}, which are of great scientific and engineering importance but are also difficult to be accurately captured by lower-order models~\cite{wu2015fidelity,WU_IHME_FUEL2016,johnson2017general} Calculations of such kind are inherently expensive due to the large number of species and reactions involved. Consequently, the efficiency of the numerical methods and their implementations becomes critical given the high overall cost at stake.

In this study, efficient strategies for the time stepping of the compressible Navier-Stokes equations with reactive scalars are considered. The drastically different characteristics and properties of the advection-diffusion and reaction parts of the equations pose great challenges to efficient treatment of the highly non-linear partial differential equations (PDEs) with a large number of degrees of freedom used to describe the evolution of turbulent reacting flows. To achieve efficient treatment of these two operators, the technique of the operator splitting, such as the Strang splitting method~\cite{strang1968construction}, is typically applied. The Strang splitting method is strongly stable and second-order accurate in time. By splitting the reaction operator from the advection-diffusion operator, different time integration methods can be applied, which can be chosen to be a better fit for the corresponding operator~\cite{wu2017application}.

For the advection-diffusion operator of the compressible Navier-Stokes equations, explicit time stepping methods are typically used in moderate-to-high Mach number applications~\cite{gottlieb2001strong}. The choice is largely motivated by the advantage of scalability in parallel calculation and the linear cost scaling with respect to the spatial resolution. The resulting time-step size is restricted by the acoustic time scale, which is inversely proportional to the grid size. For low Mach number applications, effective preconditioning schemes can be applied to eliminate the overly restrictive time-step size constraint with a similar level of computational efficiency for parallel calculations~\cite{hsieh1997preconditioned,yang2017comparison}.

For the reaction operator, its spatial independence allows the usage of implicit time stepping methods, which are more efficient in dealing with small chemical time scales that are typically associated with rapid destruction of radical species. Commonly used software packages such as CVODE~\cite{cohen1996cvode} and DASSL~\cite{petzold1982description}, adopts multi-step methods like the variable-order Backward-Differentiation Formulas (BDF)~\cite{gear1971automatic}. These methods are developed to achieve high order-of-accuracy with better efficiency compared to single-step methods for integration over a long period of time. The cost of such fully implicit integration methods scales cubicly with respect to the number of species involved. Despite the usage of reduced kinetic schemes~\cite{LU_LAW_PECS2009,yang2017global}, it can nevertheless still be overwhelmingly expensive compared to the advection-diffusion part of the governing equations~\cite{wu2016efficient}.

For complex turbulent reacting systems of interest, the aforementioned strategy can be refined to obtain improvement in both accuracy and efficiency. The conventional Strang splitting scheme, thought second-order accurate, does not preserve steady state~\cite{Speth2013}, which is shown to behave problematically for combustion systems near critical operating conditions~\cite{Lu2017}. In addition, the explicit time-step size suitable for the compressible Navier-Stokes equations of moderate-to-high Mach number flows is much smaller than the long time horizon designed for BDF-based methods. As a result, the startup issue associated with these methods makes them less preferable for the task of integrating the chemical reactions over a very short period of time. More importantly, the majority of the chemical time scales are no longer considered stiff compared with the time horizon determined by the acoustic CFL condition. Therefore, these methods are inefficient due to the fully implicit treatment of all species and reactions.

In the present study, a combination of a steady-state preserving operator splitting method and a semi-implicit integration scheme is proposed for efficient time stepping of reactive turbulent simulations with stiff chemistry. In the first part of the paper, the \textit{Simpler} balanced splitting method is constructed with reduced computational cost. The analysis shows the improved stability properties of this methods. It is then demonstrated to be capable of stable and accurate prediction of ignition and extinction for reaction-diffusion systems near critical conditions. In the second part, the ROK4E scheme is designed for semi-implicit integration of spatially independent reacting systems. Being a Rosenbrock-Krylov method, ROK4E utilizes the low-rank approximation of the Jacobian to reduce the cost for integrating the system of ODEs that have relative few stiff components. The efficiency of the scheme is further improved via the careful choice of coefficients to require three right-hand-side (RHS) evaluations over four stages. The paper is concluded with numerical examples of efficient calculation, by combing these two methods, for large-scale parallel simulations of turbulent flames.

\section{Governing Equations}
\label{S:governing_equations}
In this study, the set of advection-diffusion-reaction equations of interest is the compressible Navier-Stokes equations with $N_S$ reactive scalars. This system of equations describing the conservation of mass, momentum, total energy, and species is of the following form:
\begin{equation}
\label{e:conservation_law}
	\frac{\partial}{\partial t} \mathbf{U} = 
    \mathbf{\mathcal{T}}(\mathbf{U}) + \mathbf{\mathcal{R}}(\mathbf{U}) \,
\end{equation}
where
\begin{equation}
\label{e:conserv_varaibles}
	\mathbf{U} = 
		\begin{bmatrix}
            \rho \\ 
            \rho \mathbf{u} \\
            \rho E \\
            \rho Y_1 \\
            \vdots \\
            \rho Y_{N_S-1} \\
        \end{bmatrix} \,,
\end{equation}
\begin{equation}
\label{e:convection_diffussion}
    \mathbf{\mathcal{T}}(\mathbf{U}) = 
    	\begin{bmatrix}
            -\nabla \cdot (\rho \mathbf{u}) \\ 
            - \nabla \cdot (\rho \mathbf{u} \otimes \mathbf{u}) 
                -\nabla p + \nabla \cdot \boldsymbol{\tau} \\
            - \nabla \cdot \big[ (\rho E + p) \mathbf{u} \big] 
                -\nabla \cdot ( \mathbf{q} + \boldsymbol{\tau} \mathbf{u}) \\
            - \nabla \cdot (\rho \mathbf{u} Y_1) -\nabla \cdot \mathbf{j}_1\\
            \vdots \\
            - \nabla \cdot (\rho \mathbf{u} Y_{N_S-1}) -\nabla \cdot \mathbf{j}_{N_S-1}\\
        \end{bmatrix} \,,
\end{equation}
and
\begin{equation}
\label{e:chemical_source}
    \mathbf{\mathcal{R}}(\mathbf{U}) = 
        \begin{bmatrix}
            0 \\ 
            0 \\
            0 \\
            \dot{\omega}_1 \\
            \vdots \\
            \dot{\omega}_{N_S-1} \\
        \end{bmatrix} \,,
\end{equation}
The viscous stress tensor $\boldsymbol{\tau}$ is defined by:
\begin{equation}
	\boldsymbol{\tau} = -\frac{2}{3} \mu (\nabla \cdot \mathbf{u}) \mathbf{I} 
    	+ \mu \big[ \nabla \mathbf{u} + (\nabla \mathbf{u})^{T} \big] \, ,
\end{equation}
and the energy flux $\mathbf{q}$ is:
\begin{equation}
	\mathbf{q} = -\lambda \nabla T + \sum_{k=1}^{N_S} \mathbf{j}_k h_k ,
\end{equation}
in which the heat conduction follows Fourier's Law.
The diffusion related mass flux $\mathbf{j}_{k}$ for the $k$-th species is defined by:
\begin{equation}
\label{e:diff_flux}
	\mathbf{j}_k = 
    	- \rho D_{km} \nabla Y_k + \rho Y_k \sum_{k^{\prime}=1}^{N_S} \rho D_{k^{\prime}m} \nabla Y_{k^{\prime}} ,
\end{equation}
where the first terms is the mixture-averaged diffusion flux and the second term is the corresponding mass-conserving correction velocity.

\section{Simpler balanced splitting}
\label{S:simpler_balanced_splitting}
\subsection{Strang splitting for advection-diffusion-reaction equations}
\label{SS:strang_splitting}
For the system described in Eq.~\ref{e:conservation_law}, where the two terms, $\mathbf{\mathcal{T}}$ and $\mathbf{\mathcal{R}}$, are drastically different in their properties, the class of operator splitting methods~\cite{marchuk1968some,strang1968construction} is a natural choice to achieve efficient integration in time, such that each term can be integrated using schemes most suitable.

The transport (advection-diffusion) term, $\mathbf{\mathcal{T}}(\mathbf{U})$, represents the conservation law of mass, momentum, energy, and species in a non-reacting system. It is spatially discretized using fully-conservative schemes, through which the evaluation at one point requires the information of its neighboring points. For large-scale calculations parallelized via distributed memory systems, this also means the necessity of data transfer over network each time the function is evaluated. In addition to spatial connectivity, the transport term is also non-stiff. Despite the diffusion term, for turbulent flow simulations, the stability-limited time step size is inversely proportional to the grid size, due to the choice of resolution that is turbulent length scales for both DNS and LES. The reaction term, $\mathbf{\mathcal{R}}(\mathbf{U})$, is non-zero only for the species conservation equations. In contrast to the transport term, the reaction term is local in space, highly non-linear, and very stiff. In addition, the magnitude and stiffness of this term is often heterogeneous in space since flames are highly localized features.

The method of Strang splitting~\cite{strang1968construction} integrates the two terms sequentially in a symmetric fashion:
\begin{subequations}
\begin{align}
\label{e:strang_t1}
	d_t \mathbf{U}^{(1)} &= \mathbf{\mathcal{T}}(\mathbf{U}^{(1)}) \,, 
    	&&\mathbf{U}^{(1)}(t_n) = \mathbf{U}_n\\
\label{e:strang_r2}
	d_t \mathbf{U}^{(2)} &= \mathbf{\mathcal{R}}(\mathbf{U}^{(2)})\,, 
    	&&\mathbf{U}^{(2)}(t_n) = \mathbf{U}^{(1)}(t_n + h/2)\\
\label{e:strang_t3}
	d_t \mathbf{U}^{(3)} &= \mathbf{\mathcal{T}}(\mathbf{U}^{(3)})\,, 
    	&&\mathbf{U}^{(3)}(t_n + h/2) = \mathbf{U}^{(2)}(t_n + h) \\
	\mathbf{U}_{n+1} &= \mathbf{U}^{(3)}(t_n + h) \,.
\end{align}
\end{subequations}
The splitting formulation is second-order accurate and strongly stable. In addition, it is also symplectic for nonlinear equations~\cite{hairer2006geometric,Speth2013}.

The non-stiff steps of Eqs.~\ref{e:strang_t1} and~\ref{e:strang_t3} can be efficiently integrated using an explicit Runge-Kutta (RK) scheme, whose cost, in both time and memory, grows linearly with the degrees of freedom and the number of species. The stiff reaction step is typically  treated with an implicit integration scheme. A more efficient method for this problem, a semi-implicit Rosenbrock-Krylov scheme, will be discussed in Sec.~\ref{s:rok4e_scheme}. The spatially-decoupled reaction step of Eq.~\ref{e:strang_r2} can be solved as an assembly of 0-D reactor systems, which are constant-volume and adiabatic in this case. Various local time step sizes can be used for these systems.

It is worth noting that an alternative Strang-splitting formulation can be obtained by performing the reaction step first. The order of accuracy and stability properties are not affected by such rearrangement. However, this formulation is computationally more expensive, since the implicit/semi-implicit integration process for the stiff reaction steps is performed twice with half the step size.

\subsection{Steady-state preservation for splitting methods}
In addition to the order of accuracy and stability, steady-state preservation is another important criterion for numerically integrating a system of ordinary differential equations (ODEs). 
\theoremstyle{definition}
\begin{definition}[steady-state preserving]
\label{d:steady_state_preserving}
For a system of ODEs in the form of $$d_t \mathbf{u} = \mathbf{f}(\mathbf{u}) \,,$$
the numerical integration method is steady-state preserving, if given $\mathbf{u}_{n} = \mathbf{u}_\infty$ such that $\mathbf{f}(\mathbf{u}_\infty) = 0$, the solution of the next step remains to be $\mathbf{u}_\infty$, regardless of the step size, i.e.
\begin{equation}
	\forall h > 0,\, \mathbf{u}_{n+1} = \mathbf{u}_{n} = \mathbf{u}_\infty \,.
\end{equation}
\end{definition}
The Strang splitting formulation is second-order accurate and strongly stable but not steady-state preserving. Consider a system of linear ODEs:
\begin{equation}
\label{e:ode_linear}
	d_t \mathbf{u} = \mathbf{A} \mathbf{u} + \mathbf{a} + \mathbf{B} \mathbf{u} + \mathbf{b} \,.
\end{equation}
The system has a steady-state solution $\mathbf{u}_{\infty} = -(\mathbf{A} + \mathbf{B})^{-1} (\mathbf{a} + \mathbf{b})$. The steady state of $\mathbf{u}_{\infty}$ can be shown to be not preserved by Strang splitting, with a one-step deviation of $\mathcal{O}(h^3)$. The split system has an $h$-dependent steady state of itself. The difference between the two is $\mathcal{O}(h^2)$, as analyzed Speth et~al.~\cite{Speth2013}. 

The steady-state error of Strang splitting can have a qualitatively significant impact on dynamical systems near bifurcation. An illustrative example to the credit of Lu et~al.~\cite{Lu2017} is the near-extinction/ignition behavior of a perfectly stirred reactor (PSR), in which the ordinary Strang splitting method can lead to unphysical ignition or extinction for near-limit conditions.

\subsection{Balanced splitting and its stability}
\label{ss:balanced_splitting_and_its_stability}
The class of balanced splitting methods was first proposed by Speth et~al.~\cite{Speth2013} for the construction of second-order accurate splitting formulations that are also steady-state preserving. The key ideal is to regroup the operators by adding and subtracting a constant, such that the steady state of the combined system can be preserved in each split part. For instance, the system of Eq.\ref{e:conservation_law} can be regrouped during the $n$-th step as
\begin{equation}
\label{e:conservation_law_balanced}
\begin{split}
	\frac{\partial}{\partial t} \mathbf{U} 
    	&= \mathbf{\mathcal{T}}_n^*(\mathbf{U}) + \mathbf{\mathcal{R}}_n^*(\mathbf{U}) \\
        &= \big( \mathbf{\mathcal{T}}(\mathbf{U}) + \mathbf{c}_n \big) + 
        	\big( \mathbf{\mathcal{R}}(\mathbf{U}) - \mathbf{c}_n \big) \, .
\end{split}
\end{equation}
The integration proceeds as:
\begin{subequations}
\label{e:balanced}
\begin{align}
\label{e:balanced_t1}
	d_t \mathbf{U}^{(1)} &= \mathbf{\mathcal{T}}(\mathbf{U}^{(1)}) + \mathbf{c}_n \,, 
    	&&\mathbf{U}^{(1)}(t_n) = \mathbf{U}_n\\
\label{e:balanced_r2}
	d_t \mathbf{U}^{(2)} &= \mathbf{\mathcal{R}}(\mathbf{U}^{(2)}) - \mathbf{c}_n \,, 
    	&&\mathbf{U}^{(2)}(t_n) = \mathbf{U}^{(1)}(t_n + h/2)\\
\label{e:balanced_t3}
	d_t \mathbf{U}^{(3)} &= \mathbf{\mathcal{T}}(\mathbf{U}^{(3)}) + \mathbf{c}_n \,, 
    	&&\mathbf{U}^{(3)}(t_n + h/2) = \mathbf{U}^{(2)}(t_n + h) \\
	\mathbf{U}_{n+1} &= \mathbf{U}^{(3)}(t_n + h) \,.
\end{align}
\end{subequations}
Since the symmetric formulation of Strang is followed, the balanced splitting is also second-order accurate. The stiffness of each split step is not affected by the balancing, as they are offset by a constant. Consequently, the methods used to integrate the systems for Strang splitting are still applicable for the balanced splitting. 

One choice of the balancing factor $\mathbf{c}_n$, the simple balanced splitting proposed by Speth et~al.~\cite{Speth2013}, is
\begin{equation}
	\mathbf{c}_n = 
    \frac{1}{2} \big( \mathbf{\mathcal{R}}(\mathbf{U}) - \mathbf{\mathcal{T}}(\mathbf{U}) \big) \,.
\end{equation}
It is straightforward to verify that the simple balanced splitting is steady-state preserving following Def.~\ref{d:steady_state_preserving}.

Although $\mathbf{c}_n$ is held constant during the step from $t_n$ to $t_n+h$, its value is updated at the beginning of the next step. As such, the numerical integration error can propagate between steps through not only the solution but also the balancing factor. Therefore, the stability property of the balanced splitting methods depends on the choice of $\mathbf{c}_n$. For the simple balanced splitting, Speth et~al.~\cite{Speth2013} carried out the stability analysis using on a system of linear ODEs as defined in Eq.~\ref{e:ode_linear}. Following their notation, we define
\begin{align}
	\boldsymbol{\alpha} = e^{\mathbf{A}h/2}\,, &&  
    \boldsymbol{\beta} = e^{\mathbf{B}h}\,, && 
    \mathbf{A}^* = (\boldsymbol{\alpha} - \mathbf{I}) \mathbf{A}^{-1}\,, &&  
    \mathbf{B}^* = (\boldsymbol{\beta} - \mathbf{I}) \mathbf{B}^{-1} \, .
\end{align}

The simple balanced splitting can then be represented through a recurrence relation
\begin{equation}
\label{e:balance_recur}
	\mathbf{u}_{n+1} = \mathbf{R} \mathbf{u}_{n} - \mathbf{Q} (\mathbf{a}+\mathbf{b}) ,
\end{equation}
where 
\begin{equation}
	\mathbf{R} = 
    	\boldsymbol{\alpha} \boldsymbol{\beta} \boldsymbol{\alpha} 
        + \frac{1}{2}\big[ \boldsymbol{\alpha} \mathbf{B}^* 
        	- (\boldsymbol{\alpha} \boldsymbol{\beta} 
            	+ \mathbf{I}) \mathbf{A}^*\big] (\mathbf{A} - \mathbf{B}) ,
\end{equation}
and 
\begin{equation}
	\mathbf{Q} = 
    	\frac{1}{2} \big[ \boldsymbol{\alpha} \mathbf{B}^* 
        	+ (\boldsymbol{\alpha} \boldsymbol{\beta} + \mathbf{I}) \mathbf{A}^* \big] \, .
\end{equation}
The stability of the simple balanced splitting requires that for negative-definite $\mathbf{A}$ and $\mathbf{B}$, the spectral radius for $\mathbf{R}$ has to be less than one, i.e. $\rho\big( \mathbf{R}(h) \big) < 1$. In the limit of large step size, we have
\begin{equation}
\label{e:simple_large_h}
	\lim_{h \to \infty} \mathbf{R} =  \frac{1}{2} \big( \mathbf{I} + \mathbf{A}^{-1} \mathbf{B} \big) \,,
\end{equation}
which suggests that the method of simple balancing is unstable for large step size if $\mathbf{B}$ is much stiffer than $\mathbf{A}$. 

\begin{figure}[ht]
\begin{center}
 \subfigure[\label{f:R_simple_fixed_A} Simple balanced splitting]{\includegraphics[width = 0.49\textwidth]{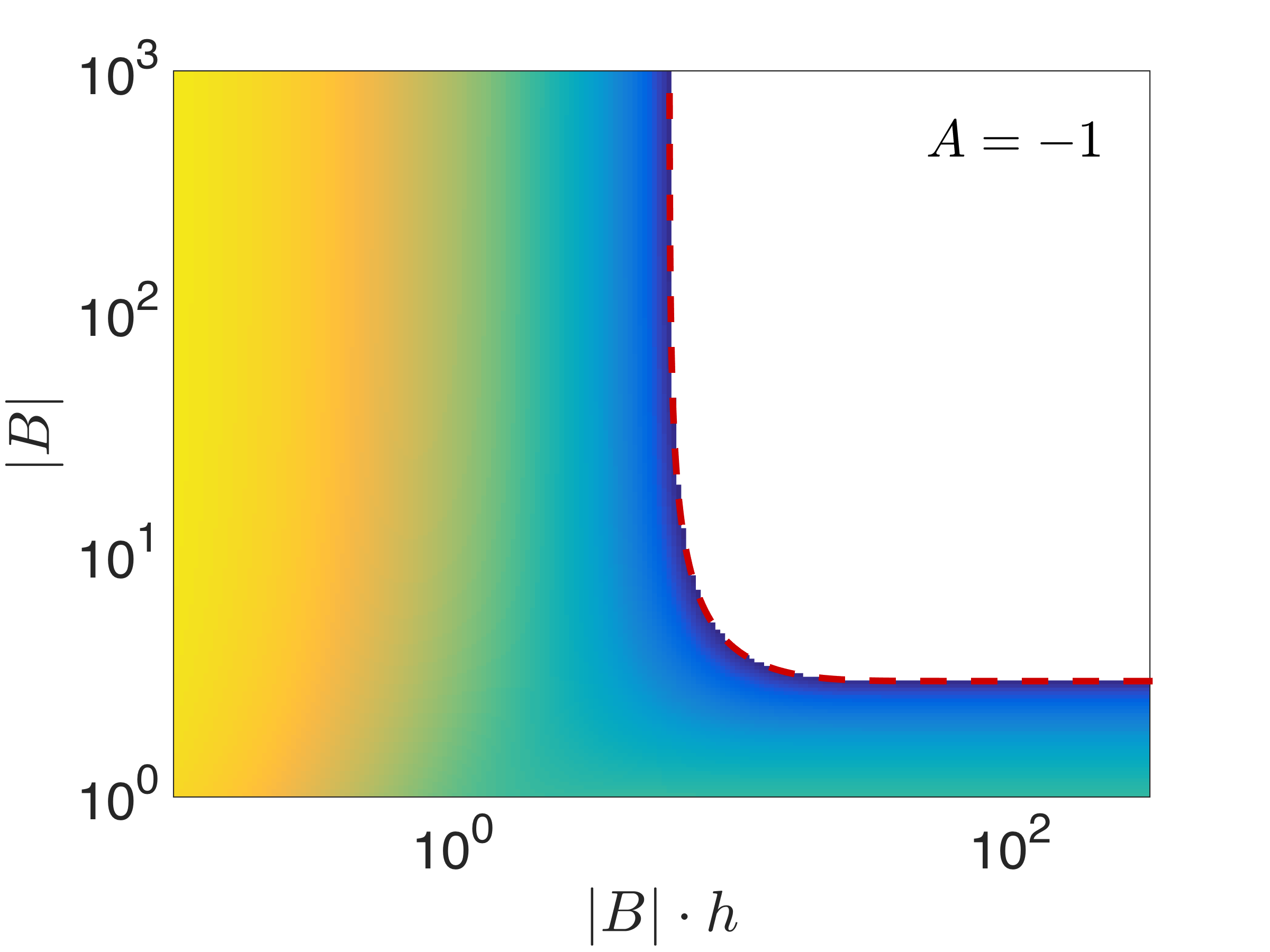}}
 \subfigure[\label{f:R_simpler_fixed_A} \textit{Simpler} balanced splitting]{\includegraphics[width = 0.49\textwidth]{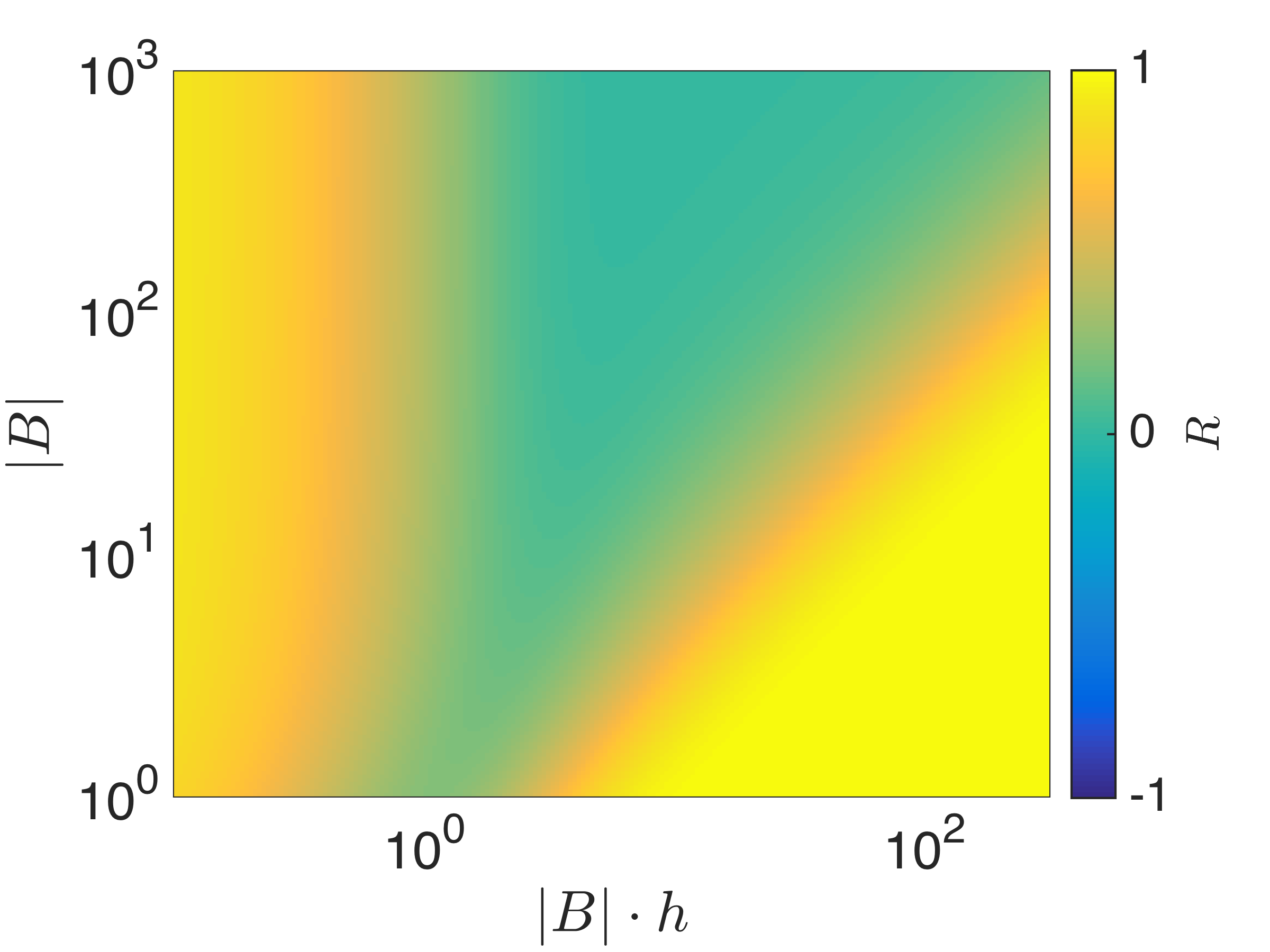}}
 \caption{\label{f:R_fixed_A} The growth factor $R$ for the scalar case as a function of $|B|$ and $|B| \cdot h$, with $A = -1$. The stability region is $R \in (-1, 1)$. The region of instability is colored in white, with the boundary highlighted by a dashed red line.}
\end{center}
\end{figure}

The restriction on the step size due the stability requirement can be well illustrated in the scalar case. As shown in Eq.~\ref{e:simple_large_h}, simple balancing is stable for all $h > 0$, only if $|B| > 3|A|$. Another limit of practical interest is when $|B| \to \infty$. In that case, the value of $R$ approaches
\begin{equation}
	\lim_{|B| \to \infty} R = \frac{1}{4}(1+\beta)(2 + B \cdot h) \,.
\end{equation}
Consequently, the method is stable for all $B < 0$, only if $|B| \cdot h < 5.99$. The contour of $R$ as a function of $|B|$ and $|B| \cdot h$ is shown in Fig.~\ref{f:R_simple_fixed_A}. Instability occurs (white region) for the scalar case if neither of the aforementioned criterion is satisfied.

The stability analysis for the scalar case suggests that, if the non-stiff system is solved first within a simple balanced splitting step, the step size is restricted to be inversely proportional to the spectral radius of the stiff operator. In that case, little benefit can be gained via operator splitting. If the stiff system is solved first, there the stability restriction can be alleviated.

\subsection{Simpler Balanced splitting}
A new balancing formulation, named \textit{simpler} balanced splitting, is proposed with improved stability properties. The \textit{simpler} balanced splitting proceeds as described in Eq.~\ref{e:balanced}, in which the non-stiff system is solved first. The balancing constant constant to be \begin{equation}
	\mathbf{c}_n = - \mathbf{\mathcal{T}}(\mathbf{U}) \,.
\end{equation}

It is straightforward to verify that the \textit{simpler} balanced splitting is also second-order accurate and satisfies the steady-state preserving conditions of Def.~\ref{d:steady_state_preserving}. Following the same linear analysis in Sec.~\ref{ss:balanced_splitting_and_its_stability}, the method of \textit{simpler} balancing can be expressed in the recurrence form as in Eq.~\ref{e:balance_recur}., with

\begin{equation}
	\mathbf{R} = 
    	\boldsymbol{\alpha} \boldsymbol{\beta} \boldsymbol{\alpha} 
        + \big[ \boldsymbol{\alpha} \mathbf{B}^* 
        	- (\boldsymbol{\alpha} \boldsymbol{\beta} 
            	+ \mathbf{I}) \mathbf{A}^*\big] \mathbf{A} ,
\end{equation}
and 
\begin{equation}
	\mathbf{Q} = \boldsymbol{\alpha} \mathbf{B}^* \, .
\end{equation}
In the limit of large step size, we have
\begin{equation}
\label{e:simpler_large_h}
	\lim_{h \to \infty} \mathbf{R} = \mathbf{I} \,.
\end{equation}
The method is stable for all $h > 0$, although the damping becomes negligible if both operators are very stiff. If only one operator becomes very stiff, i.e. in the limit of $|\mathbf{B}| \to \infty$, we have
\begin{equation}
	\lim_{|\mathbf{B}| \to \infty} \mathbf{R} = 
    	\mathbf{I} - \boldsymbol{\alpha}^2 \,.
\end{equation}
In this case, a non-negligible amount of damping is applied for $A < 0$. The contour of $|\mathbf{R}|$ in the scalar case is shown in Fig.~\ref{f:R_simpler_fixed_A} for comparison.

Due to the choice of the balancing constant, the first sub-step of \textit{simpler} balanced splitting is always at equilibrium and thus can be omitted. Combined with not having to evaluate the stiff component for the balancing constant, the \textit{simpler} balancing formulation is computationally more efficient by reducing the number of RHS evaluations. For instance, if the non-stiff system is integrated using a three-stage RK method, the \textit{simpler} balancing needs two less evaluations of the non-stiff component and one less for the stiff part compared to the simple balanced formulation. The resulting second-order two step splitting method is
\begin{subequations}
\label{e:simpler_balanced}
\begin{align}
\mathbf{c}_n &= -\mathbf{\mathcal{T}}(\mathbf{U}_n) \\
\label{e:simpler_balanced_r1}
	d_t \mathbf{U}^{(1)} &= \mathbf{\mathcal{R}}(\mathbf{U}^{(1)}) - \mathbf{c}_n \,, 
    	&&\mathbf{U}^{(1)}(t_n) = \mathbf{U}_n\\
\label{e:simpler_balanced_t2}
	d_t \mathbf{U}^{(2)} &= \mathbf{\mathcal{T}}(\mathbf{U}^{(2)}) + \mathbf{c}_n \,, 
    	&&\mathbf{U}^{(2)}(t_n + h/2) = \mathbf{U}^{(1)}(t_n + h)\\
	\mathbf{U}_{n+1} &= \mathbf{U}^{(2)}(t_n + h) \,.
\end{align}
\end{subequations}

\subsection{Numerical examples}
\label{ss:simpler_test_cases}

The effectiveness of the \textit{simpler} balanced splitting method is demonstrated via two test cases featuring perfectly stirred reactors (PSRs), both of which are adapted from the work of Lu et~al.~\cite{Lu2017}.

The first case is a non-dimensional unsteady model PSR, described by a scalar ODE as
\begin{equation}
\label{e:scalar_psr}
	d_{\tilde{t}} \widetilde{T} = 
    (\widetilde{T}_{ad} - \widetilde{T})\exp(-\widetilde{T}_a / \widetilde{T}) 
    + \frac{1}{Da} (\widetilde{T}_{in} - \widetilde{T}) ,
\end{equation}
where $\widetilde{T}_{in} = 0.15$, $\widetilde{T}_{ad} = 1.15$, and $\widetilde{T}_{a} = 1.8$. The S-curve of this system has two turning points at $Da_E = 15.90$ and $Da_I = 832.84$ for extinction and ignition respectively. Tow near-limit operating conditions are chosen for this case: 
\begin{itemize}
\item[(a)] $Da = 15.89$ and $\widetilde{T}(0) = 1.00$ for extinction;
\item[(b)] $Da = 833.0$ and $\widetilde{T}(0) = 0.15$ for ignition.
\end{itemize}

\begin{figure}[ht]
\begin{center}
 \subfigure[\label{f:psr_scalar_ext} Near-limit extinction]{\includegraphics[width = 0.49\textwidth]{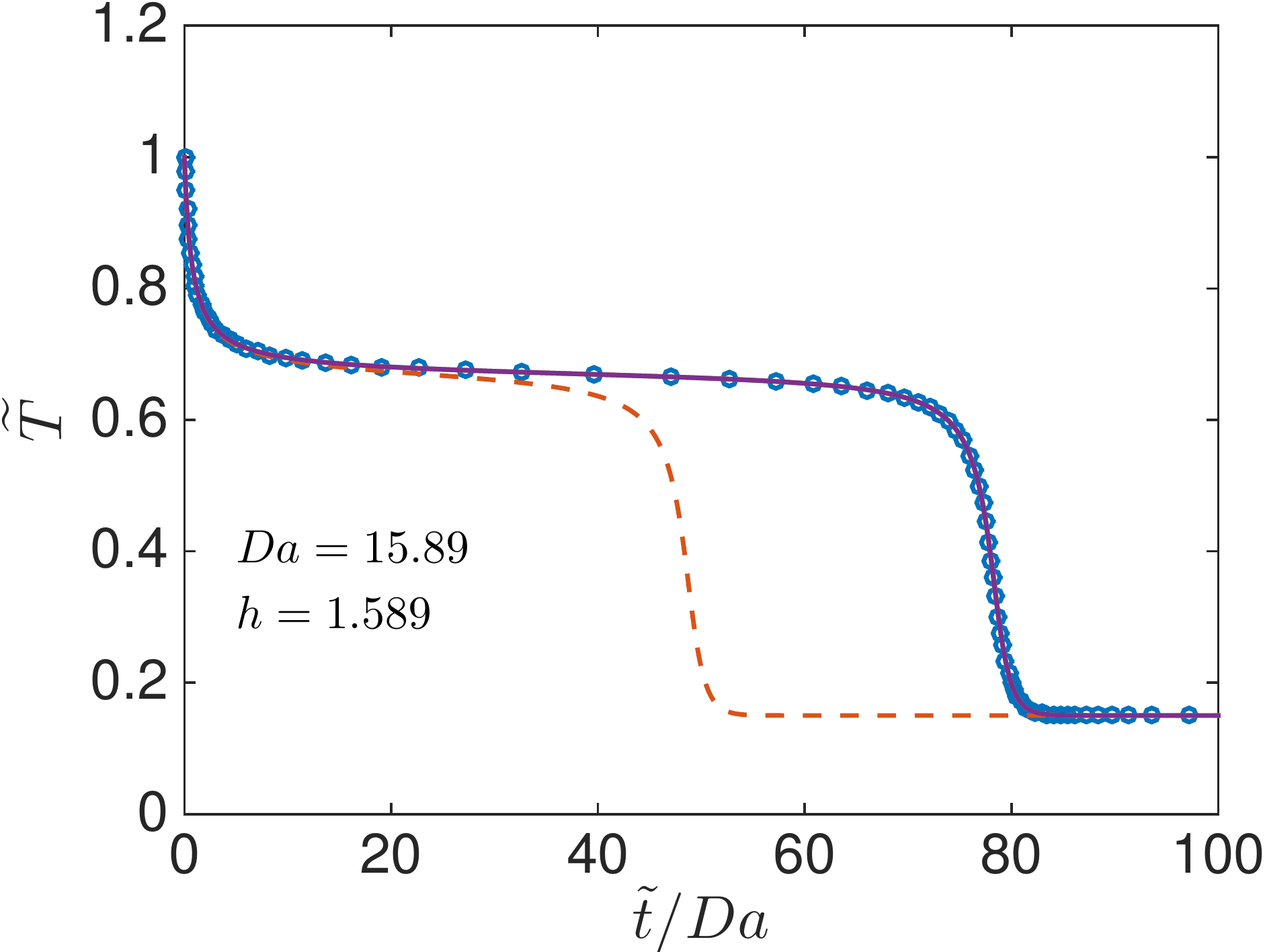}}
 \subfigure[\label{f:psr_scalar_ign} \textit{Simpler} Near-limit ignition]{\includegraphics[width = 0.49\textwidth]{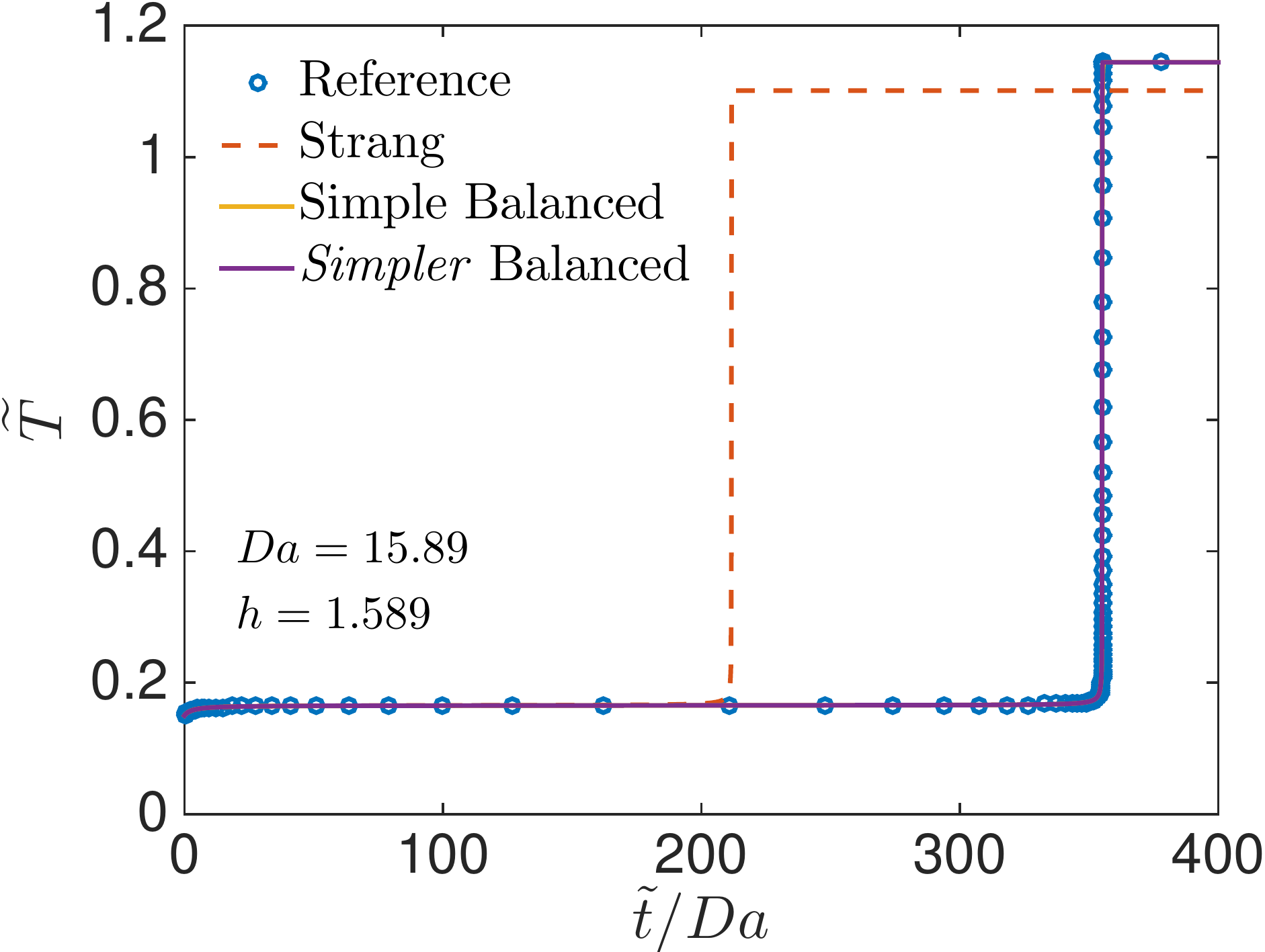}}
 \caption{\label{f:psr_scalar}
 The near-limit extinction and ignition process for the scalar unsteady PSR of Eq.~\ref{e:scalar_psr}. The reference results (blue circles) are sampled every 25 solution points.}
\end{center}
\end{figure}

The results of ordinary Strang splitting, simple balanced splitting, and \textit{simpler} balanced splitting are compared against the referenced solution, which is obtained from integrating the whole system using \texttt{ode15s}~\cite{shampine1997matlab}. A constant step size of $h = Da / h$ is chosen for all the splitting results. Within each splitting step, the mixing and the reaction parts are integrated using \texttt{ode45} and \texttt{ode15s}~\cite{shampine1997matlab} respectively. The relative and absolute tolerances for all cases are set to be $10^{-12}$ and $10^{-16}$ respectively to isolate the splitting error. The reaction step is first calculated for the simple balanced splitting, while the simpler balancing and ordinary Strang methods take the mixing step first.

The comparisons are shown in Fig.~\ref{f:psr_scalar}. Both the simple and \textit{simpler} balancing gives the correct steady state solutions and accurate prediction of the near-limit behaviors. The ordinary Strang splitting results in an early onset of the ignition/extinction event as analyzed by Lu et~al.~\cite{Lu2017}. No visible difference can be observed between the results obtained by the two balanced splitting methods.

\begin{figure}[ht]
\begin{center}
 \subfigure[\label{f:psr_h2air_ign} Temperature profile]{\includegraphics[width = 0.49\textwidth]{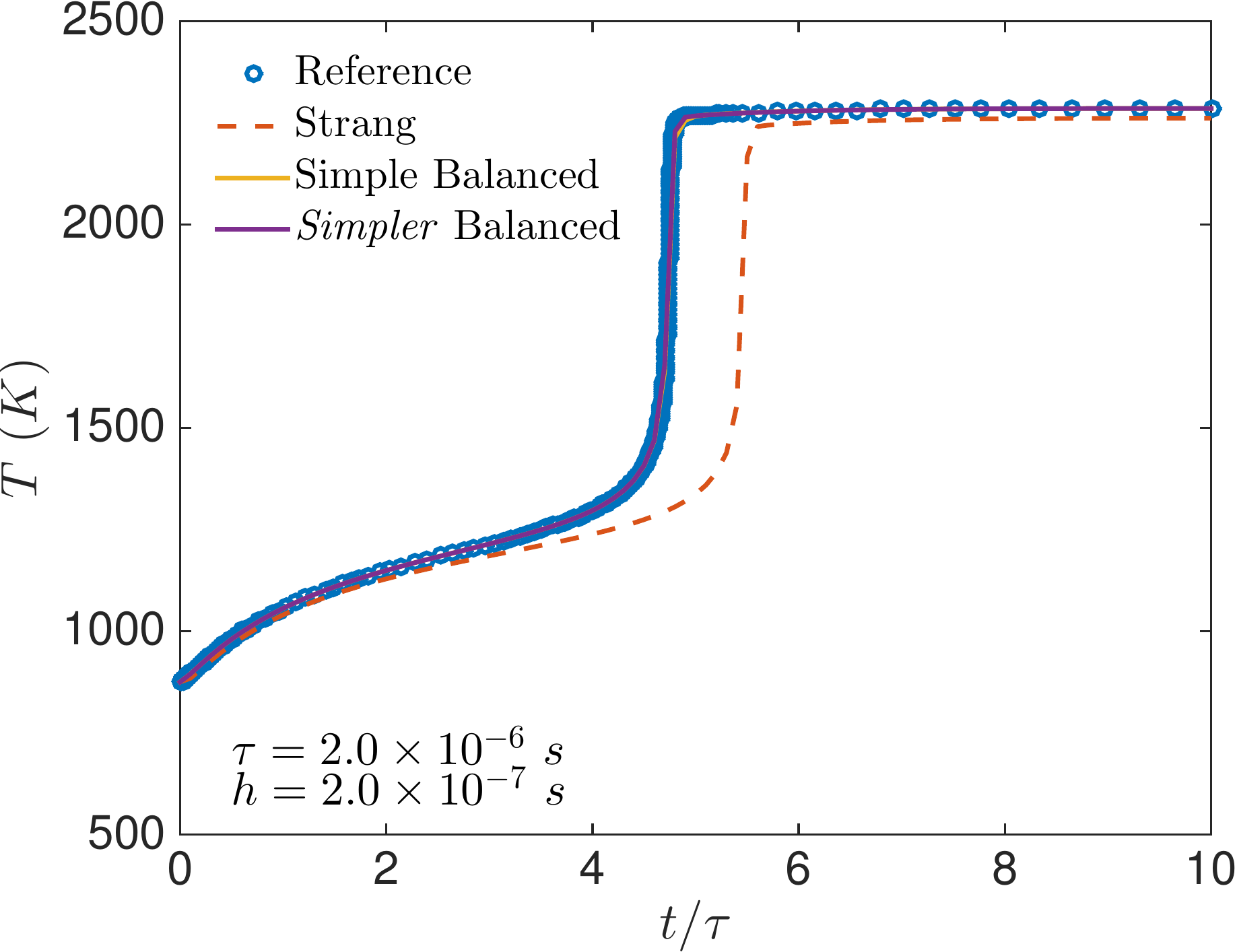}}
 \subfigure[\label{f:psr_h2air_err} Relative error of temperature]{\includegraphics[width = 0.49\textwidth]{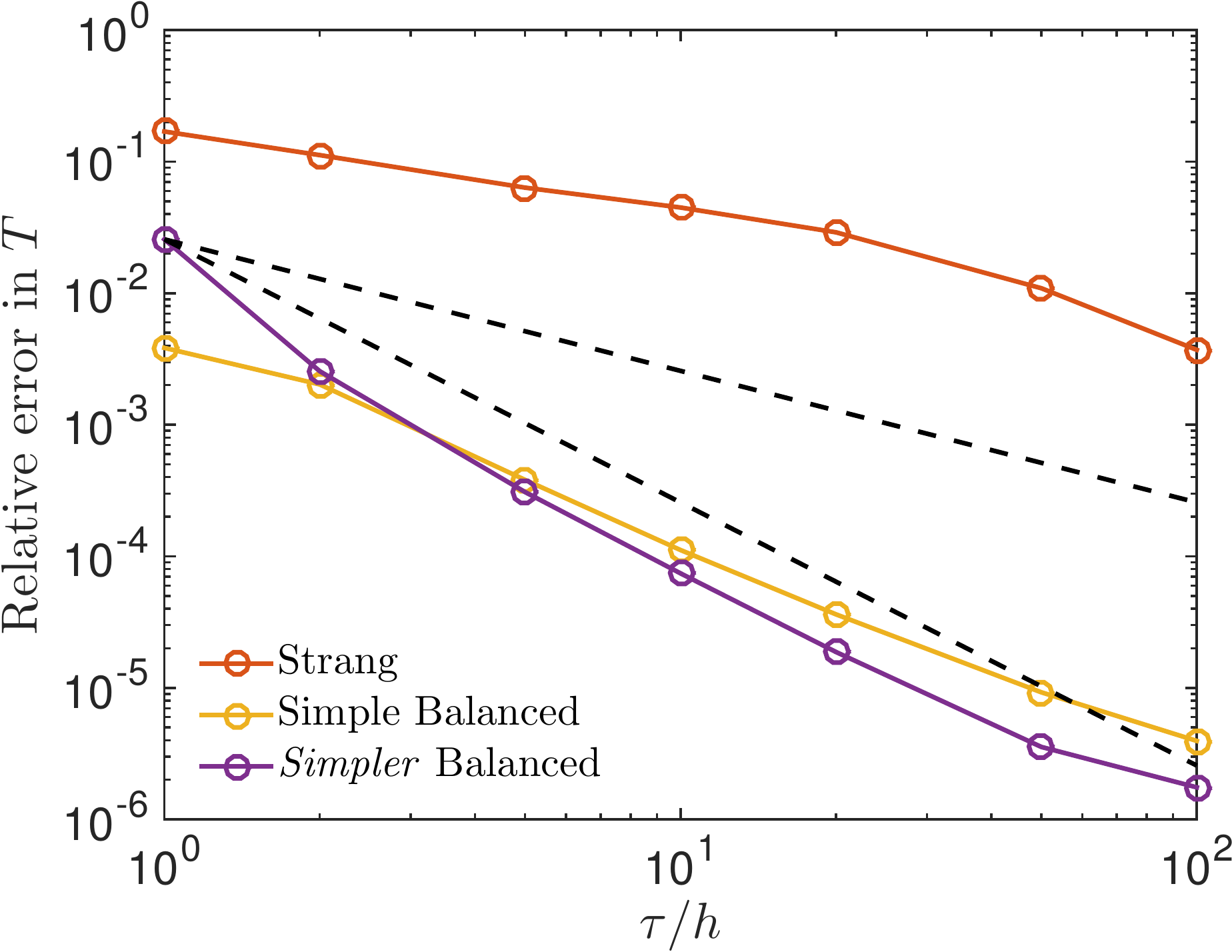}}
 \caption{\label{f:psr_h2air}
The ignition process for an unsteady PSR of Eq.~\ref{e:multi_psr}, with $\ce{H_2}$/air of $\Phi = 0.5$. The inlet is set to be the fresh mixture with an additional $0.1\%$ $\ce{H}$ radical enrichment in mass. (a): The evolution of temperature, with the reference results (blue circles) are sampled every 10 solution points. (b): The relative error in temperature at $t / \tau = 4$. The slopes of the first and second order convergence are shown in black dashed lines.}
\end{center}
\end{figure}

The second case features the unsteady PSR with a hydrogen mechanism of Burke at~al.~\cite{BURKE_CHAOS_JU_DRYER_KLIPPENSTEIN_2011}. The system of ODEs involves species and temperature equations in the form of
\begin{subequations}
\label{e:multi_psr}
\begin{align}
	d_t Y_k &= \frac{1}{\tau} (Y_{k,\,in} - Y_k) + \frac{1}{\rho}\dot{\omega}_{k} \\
    c_p d_t T &= \frac{1}{\tau} \sum_{k}^{N_S}Y_{k,\,in}(h_{k,\,in} - h_{k}) 
    	- \frac{1}{\rho} \sum_{k}^{N_S} h_k \dot{\omega}_{k} \, .
\end{align}
\end{subequations}
The ignition process of a hydrogen/air mixture with $\Phi = 0.5$ at $p = 80\, \text{atm}$ and $T = 875\, \text{K}$ is considered. The inlet stream is set to be the fresh mixture, with an additional $0.1\%$ $\ce{H}$ radical enrichment in mass. The enrichment of $\ce{H}$ radical was shown in literature~\cite{Gao2015} to cause inaccurate prediction by the ordinary Strang splitting. The residence time is set to be $\tau = 2.0 \time 10^{-6} $ $s$ as in~\cite{Lu2017}. The methods of integration are the same as in the scalar PSR case. As shown in Fig.~\ref{f:psr_h2air_ign}, the ordinary Strang splitting predicts significantly delayed ignition. Improvement in the ignition-delay time is achieved via the balanced splitting methods. A quantitative comparison between the three methods are is plotted in Fig.~\ref{f:psr_h2air_err}. The relative error of temperature is measured at $t / \tau = 4$. The \textit{simper} balancing achieved smallest error at all but largest step sizes. Both balanced formulations exhibit second-order convergence, while the ordinary Strang splitting shows slower convergence until $\tau / h = 50$.



\section{ROK4E: a forth-order Rosenbrock-Krylov scheme}
\label{s:rok4e_scheme}
\subsection{Rosenbrock-Krylov methods}
The chemical reacting system obtained from operator-splitting methods has a time horizon limited by the slitting step-size. The step-size of the splitting method is chosen to be twice of $h_{\text{CFL}}$. For high-fidelity simulations of turbulent flows, splitting step-size is typically very small due to the fine grid and high flow-speed. Thus, fewer time steps need to be taken within the horizon of the chemical system and fewer components are considered stiff. These trends favors the use of semi-implicit one-step integrators as opposed to fully-implicit multi-step method, such as the variable-order BDF methods commonly used in popular ODE solvers~\cite{cohen1996cvode,brown1994vodpk}.

From these observations, we choose the Rosenbrock-Krylov methods in this study. Proposed by Tranquilli and Sandu~\cite{Tranquilli2014}, they are a class of Rosenbrock-type integrators based on a Krylov space solution of linear systems. It features the benefit of ROW-methods~\cite{kaps1979generalized} being one-step and non-iterative while avoid the requirement of obtaining and solving the full Jacobian system at each time step. This leads to a different approach to efficient integration of the chemical reacting systems compared with the semi-implicit methods proposed for simulations of turbulent flames in the literature~\cite{savard2015computationally,yang2017parallel,macart2016semi}.

For an autonomous problem of dimension $N$, an $s$-stage Rosenbrock method~\cite{hairer1996solving} is given by 
\begin{subequations}
\label{e:s_stage_row}
\begin{align}
	(\mathbf{I} - h \gamma_{ii} \mathbf{A}_n)\mathbf{k}_i &= 
    	\mathbf{f} \bigg( \mathbf{u}_n + h  \sum_{j=1}^{i-1} \alpha_{ij}\mathbf{k_j} \bigg) 
        + \mathbf{A}_n \sum_{j=1}^{i-1} \gamma_{ij} \mathbf{k}_j, \quad i = 1,\cdots,s \\
    \mathbf{u}_{n+1} &= \mathbf{u}_{n} + h \sum_{j=1}^{s} b_j \mathbf{k}_j \,.
\end{align}
\end{subequations}
Note that the matrix $\mathbf{A}_n$ is kept the same through out the $s$ stages. Therefore, if we further require that
\begin{equation}
	 \gamma_{ii} = \gamma, \quad i = 1,\cdots,s
\end{equation}
the matrix factorization can be performed only once every time step although as much as $s$ linear systems are solved. 

In classical ROW-methods, the matrix $\mathbf{A}_n$ is chosen to be the Jacobian of $\mathbf{f}$ evaluated at $t_n$: 
\begin{equation}
\label{e:exact_jac_cond}
	\mathbf{A}_n = \mathbf{J}_n = \partial_{\mathbf{u}} \mathbf{f} (\mathbf{u}_n) \,.
\end{equation}
Methods of order 4 can be formulated with eight order conditions that can be satisfied in four stages. Many popular schemes of this type~\cite{hairer1996solving} also features embedded third-order error estimator to adaptive step-size control. Fortran implementations of these schemes can be found in the \texttt{RODAS} and \texttt{ROS4} codes~\cite{hairer1996solving}.

Being one-step, the ROW-methods avoid the issue of restart which can be problematic for turbulent combustion simulations if multi-step integrators are used. However, its competitiveness in comparison to multi-step methods like BDF is weakened by the explicit presence of the exact Jacobian in the formulation, which prohibits the application of Jacobian reuse and matrix-free techniques which are often employed by implicit multi-step solvers~\cite{cohen1996cvode,brown1994vodpk}. A class of matrix-free Rosenbrock-type integrators, named Krylov-ROW, was first proposed in the 1990s~\cite{weiner1997rowmap}, in which the Jocobian matrix was approximated via a multiple Arnoldi process~\cite{schmitt1995matrix}. More recently, a new family of integrators, called  Rosenbrock–Krylov, is developed by Tranquilli and Sandu~\cite{Tranquilli2014}. The incorporation of the Krylov-subspace properties into the order-condition theory gives methods requiring only a single Arnoldi procedure and minimal number of order conditions in addition to those of the classical ROW-methods.

For Rosenbrock-Krylov methods of order $p$, the exact Jacobian condition of Eq.~\ref{e:exact_jac_cond} is replaced by
\begin{equation}
\label{e:krylov_jac_cond}
	\mathbf{A}_n^k \mathbf{f}_n = \mathbf{J}_n^k \mathbf{f}_n, \quad 0 \le k \le p-1 \,,
\end{equation}
where $\mathbf{f}_n = \mathbf{f}(\mathbf{u}_n)$. This condition can be satisfied by an approximation of $\mathbf{J}_n$ in the Krylov-subspace of dimension $M \ge p$. The Krylov-subspace is chosen to be
\begin{equation}
	\mathcal{K}_M(\mathbf{J}_n,\, \mathbf{f}_n) = 
    	\text{span} \big\{ \mathbf{f}_n,\, \mathbf{J}_n\mathbf{f}_n,\, \cdots,\, \mathbf{J}_n^{M-1}\mathbf{f}_n \big\} \,.
\end{equation}
The corresponding Arnoldi iteration~\cite[Lec.~33]{trefethen1997numerical} gives an orthonormal matrix $\mathbf{Q}_{n} \in \mathbb{R}^{N \times M}$ and an upper Hessenberg matrix $\mathbf{H}_{n} \in \mathbb{R}^{M \times M}$ such that
\begin{equation}
	\mathbf{H}_{n} = \mathbf{Q}_{n}^T \mathbf{J}_n \mathbf{Q}_{n} \,.
\end{equation}
The Jacobian is approximated by its projection to $\mathcal{K}_M(\mathbf{J}_n,\, \mathbf{f}_n)$ as
\begin{equation}
\label{e:jac_proj}
	\mathbf{A}_{n} = \mathbf{Q}_{n} \mathbf{H}_n \mathbf{Q}_{n}^T 
    	= \mathbf{Q}_{n} \mathbf{Q}_{n}^T \mathbf{J}_n \mathbf{Q}_{n} \mathbf{Q}_{n}^T\,.
\end{equation}
It is straightforward to verify that $\mathbf{A}_{n}$, the rank-$M$ approximation of $\mathbf{J}_{n}$, satisfies the condition of Eq.~\ref{e:krylov_jac_cond} for $M \ge p$. In addition, the exact Jacobian is recovered, i.e.~$\mathbf{A}_{n} = \mathbf{J}_{n}$, when $M = N$.

A four-stage, fourth-order Rosenbrock-Krylov scheme can be constructed with $9$ order conditions as follows:
\begin{subequations}
\label{e:order_cond_ros_k}
\begin{align}
& b_1 + b_2 + b_3 + b_4 = 1 \,,\\
& b_2 \beta^\prime_2 + b_3 \beta^\prime_3 + b_4 \beta^\prime_4 = 1/2 - \gamma \,,\\
& b_2 \alpha^2_2 + b_3 \alpha^2_3 + b_4 \alpha^2_4 = 1/3 \,,\\
& b_3 ( \beta_{32} \beta_2^\prime ) + b_4( \beta_{42}\beta_2^\prime + \beta_{43}\beta_3^\prime) 
	= 1/6 - \gamma + \gamma^2 \,,\\
& b_2 \alpha^3_2 + b_3 \alpha^3_3 + b_4 \alpha^3_4 = 1/4 \,,\\
& b_3 \alpha_3 \alpha_{32} \beta_2^\prime + b_4 \alpha_4(\alpha_{42} \beta_2^\prime + \alpha_{43} \beta_3^\prime) 
	= 1/8 - \gamma/3 \,,\\
\label{e:order_cond_special_1}
& b_3\alpha_{32}\alpha_2^2 + b_4(\alpha_{42}\alpha_2^2 + \alpha_{43}\alpha_3^2) = 1/12\\
\label{e:order_cond_special_2}
& b_3\gamma_{32}\alpha_2^2 + b_4(\gamma_{42}\alpha_2^2 + \gamma_{43}\alpha_3^2) = -\gamma/3 \\
& b_4\beta_{43}\beta_{32}b_2^\prime = 1/24 - \gamma/2 + 3\gamma^2/2 - \gamma^3 \,,
\end{align}
\end{subequations}
in which the following abbreviations are used following the conventions in literature
\begin{equation}
	\beta_{ij} = \alpha_{ij} + \gamma_{ij} \,, \quad 
    \alpha_{i} = \sum_{j=1}^{i-1} \alpha_{ij} \,, \quad 
    \beta^\prime_{i} = \sum_{j=1}^{i-1} \beta_{ij} \,.
\end{equation}
It is worth noting that the order conditions in Eq.~\ref{e:order_cond_ros_k} are identical to those of the classical forth-order ROW-methods, except for those in Eq.~\ref{e:order_cond_special_1} and~\ref{e:order_cond_special_2}, which split the $t_{43}$ condition of ROW-methods into two separate ones.

In the case of exact Jacobian, i.e.~$\mathbf{A}_n = \mathbf{J}_n$, the stability function for the class of Rosenbrock-Krylov methods is in the form of
\begin{equation}
\label{e:stability_exact_jac}
	R(z) = 1 + z \mathbf{b}^T(\mathbf{I} - z\boldsymbol{\beta})^{-1} \mathbbm{1} \,,
\end{equation}
where $\mathbf{b}^T, (b_1,\, \cdots ,\, b_s)$ and $\boldsymbol{\beta} = (\beta_{ij})_{i,\,j = 1}^{s}$, and $\mathbbm{1} \in \mathbb{R}^s$ is a vector of ones. This stability function is equal to that of the ROW-methods and DIRK-methods~\cite{alexander1977diagonally}.

\subsection{ROK4E: An efficient four-stage, fourth-order, L-stable method}

A four-stage, fourth-order method in the form of Eq.~\ref{e:s_stage_row} has $17$ parameters, which need to be chosen to satisfy $9$ order conditions as stated in Eq.~\ref{e:order_cond_ros_k}. Thus, there leaves $8$ degrees of freedom to construct the schemes with additional desirable properties.

For the ROK4E method, we first choose $\gamma = 0.573$ such that $R(\infty) = 0$ for the stability function in Eq.~\ref{e:stability_exact_jac}, which gives L-stability in the case of exact Jacobian. It is further required that 
\begin{equation}
	\alpha_{43} = 0, \quad \alpha_{42} = \alpha_{32}, \quad \alpha_{41} = \alpha_{31} \,,
\end{equation}
which make the argument of $\mathbf{f}$ in Eq.~\ref{e:s_stage_row} be the same for $i=3$ and $i=4$. Hence, the number of function evaluations is reduced by one. Similar to the construction of Kaps-Rentrop~\cite{kaps1979generalized}, it is further required that two additional fifth order conditions ($t_{51}$ and $t_{56}$) are also satisfied. For the sake of step size control, an embedded formula of order $3$ is also needed. The embedded method of order $3$ is in the form of
\begin{equation}
\widehat{\mathbf{u}}_{n+1} = \mathbf{u}_{n} + h \sum_{j=1}^{s} \hat{b}_j \mathbf{k}_j,
\end{equation}
which uses the same $\mathbf{k}_j$ but different weight coefficients $\hat{b}_j$. This can be achieved by making the linear system of conditions (a)-(d) in Eq.~\ref{e:order_cond_ros_k} singular.

The additional $7$ conditioned in combination with the $9$ order conditions leaves only one degree of freedom in determining the $17$ coefficients of the scheme. The choice of $b_3 = 0$ is made, which leads to the coefficient values for this methods, named ROK4E, as given in.

\begin{table}[ht!]
\centering
\begin{tabular}{l c r l c r} 
 \hline
 $\gamma$ & $=$ & $0.572816062482135$ &  &  &  \\
 $\gamma_{21}$ & $=$ & $-0.602765307997356$ &  &  &  \\
 $\gamma_{31}$ & $=$ & $-1.389195789724843$ & $\gamma_{32}$ & $=$ & $1.072950969011413$ \\ 
 $\gamma_{41}$ & $=$ & $0.992356412977094$ & $\gamma_{42}$ & $=$ & $-1.390032613873701$ \\ 
 $\gamma_{43}$ & $=$ & $-0.440875890223325$ &  &  &  \\ 
 \hline
 $\alpha_{21}$ & $=$ & $0.432364435748567$ &  &  &  \\
 $\alpha_{31}$ & $=$ & $-0.514211316876170$ & $\alpha_{32}$ & $=$ & $1.382271144617360$ \\ 
 $\alpha_{41}$ & $=$ & $-0.514211316876170$ & $\alpha_{42}$ & $=$ & $1.382271144617360$ \\ 
 $\alpha_{43}$ & $=$ & $0$ &  &  &  \\ 
 \hline
 $b_{1}$ & $=$ & $0.194335256262729$ & $b_{2}$ & $=$ & $0.483167813989227$ \\ 
 $b_{2}$ & $=$ & $0$ & $b_{3}$ & $=$ & $0.322496929748044$ \\ 
 \hline
 $\hat{b}_{1}$ & $=$ & $-0.217819895945721$ & $\hat{b}_{2}$ & $=$ & $1.03130847478467$ \\ 
 $\hat{b}_{2}$ & $=$ & $0.186511421161047$ & $\hat{b}_{3}$ & $=$ & $0$ \\ 
 \hline
\end{tabular}
\label{t:coeff_rok4e}
\caption{Coefficients of ROK4E}
\end{table}

\subsection{Implementation}
The Rosenbrock-type methods, if implemented directly using the form of Eq.~\ref{e:s_stage_row}, requires a matrix-vector multiplication of $\mathbf{A}_n$ in addition to solving a linear system of $\mathbf{I}-h\gamma\mathbf{A}_n$. The matrix-vector multiplication can be avoided by transforming Eq.~\ref{e:s_stage_row} into the following form:
\begin{equation}
\label{e:s_stage_row_transformed}
	(\mathbf{I} - h\gamma\mathbf{A}_n ) 
    	(\mathbf{k}_i + \sum_{j=1}^{i-1} \frac{\gamma_{ij}}{\gamma} \mathbf{k}_j) =
        \mathbf{f} \bigg( \mathbf{u}_n + h \sum_{j=1}^{i-1} \alpha_{ij}\mathbf{k}_j \bigg) 
        + \sum_{j=1}^{i-1} \frac{\gamma_{ij}}{\gamma} \mathbf{k}_j.
\end{equation}

The orthonormal matrix $\mathbf{Q}_{n} \in \mathbb{R}^{N \times M}$ and an upper Hessenberg matrix $\mathbf{H}_{n} \in \mathbb{R}^{M \times M}$, which are used to approximate $\mathbf{J}_m$ as in Eq.~\ref{e:jac_proj}, are obtained from the Arnoldi iteration.The Jacobian-vector products in the Arnoldi process can be approximated matrix-free, as in Newton-Krylov methods~\cite{knoll2004jacobian}, by finite difference of the form 
\begin{equation}
	\mathbf{J}_n \mathbf{b} \approx \big(\mathbf{f}(\mathbf{u}_n + \delta \mathbf{b}) 
    	- \mathbf{f}(\mathbf{u}_n) \big) / \delta \, .
\end{equation}
Using the matrix-free approximation, $\mathbf{Q}_{n}$ and $\mathbf{H}_{n}$ of can be obtained via the Arnoldi iteration with $M$ evaluations of $\mathbf{f}(\cdot)$.

The linear system of $(\mathbf{I} - h\gamma \mathbf{A}_n)$ can be efficiently solved by exploiting the low rank nature of $\mathbf{A}_n = \mathbf{Q}_{n} \mathbf{H}_n \mathbf{Q}_{n}^T$. The inversion can be obtained by using twice the Woodbury matrix identity~\cite{woodbury1950inverting}:
\begin{equation}
\label{e:solve_kry_jac}
\begin{aligned}
	(\mathbf{I} - h\gamma \mathbf{A}_n)^{-1} &= (\mathbf{I} - h\gamma \mathbf{Q}_{n} \mathbf{H}_n \mathbf{Q}_{n}^T)^{-1} \\
    &= \mathbf{I} - \mathbf{Q}_{n}(\mathbf{I} - \frac{1}{h\gamma} \mathbf{H}_n^{-1})^{-1}\mathbf{Q}_{n}^T \\
    &= \mathbf{I} - \mathbf{Q}_{n}\big( \mathbf{I} - (\mathbf{I} - h\gamma \mathbf{H}_n)^{-1}\big)\mathbf{Q}_{n}^T \,.
\end{aligned}
\end{equation}
As such, a linear system of size $M \times M$ needs to be solved instead of the full system of size $N \times N$.

Adaptive step-size control is achieved via the estimation of error using the embedded scheme. The normalized error is estimated to be
\begin{equation}
	err_{n+1} = |(\widehat{\mathbf{u}}_{n+1} - \mathbf{u}_{n+1}) / (Rtol \cdot \mathbf{u}_{n+1} + Atol)| \,,
\end{equation}
where $Rtol$ and $Atol$ are the user-specified relative and absolute tolerance respectively. Since the Rosenbrock-Krylov methods are semi-implicit, the system may be unstable if excessively large step-size is used. The adaptive step-size control can avoid the unstable region by reducing the step-size once the estimated error is too large. The method of PI step-size control is used in this study, which improves the step-control stability. This avoids spurious oscillations of the step-size and the frequent rejection due to large error. The step-size is updated using the error estimation at both current and previous steps as~\cite{gustafsson1991control}
\begin{equation}
h_{n+1} = h_n \cdot \min\big\{5.0,\, \max\{0.2,\, 0.8 \cdot err_{n}^{\beta} / err_{n+1}^{\alpha} \big\} \big\} ,\,
\end{equation}
where $\alpha = 0.7/p$, $\beta = 0.4/p$, and $p = 4$ for ROK4E. The pseudo code for the ROK4E integrator is outlined in Alg.~\ref{a:ROK4E}.

\begin{algorithm}[!ht]
\caption{ROK4E integrator for an autonomous system}
\label{a:ROK4E}
\begin{algorithmic}[1]
\Procedure{ROK4E}{$t_{\text{end}}$, $h_0$}
\State $t \gets 0$, $n \gets 0$
\While{$t < t_{\text{end}}$}
	\State $\mathbf{H}_n ,\, \mathbf{Q}_n \gets \text{Arnoldi}(\mathbf{f}_n,\, M)$
	\For{$i = 1, \cdots , s$}
    	\State $\mathbf{F}_i = \mathbf{f} \bigg( \mathbf{u}_n + h_n \sum_{j=1}^{i-1} \alpha_{ij}\mathbf{k}_j \bigg) 
        	+ \sum_{j=1}^{i-1} \frac{\gamma_{ij}}{\gamma} \mathbf{k}_j$ 
        \State $\mathbf{k}_i = \mathbf{F}_i 
        	- \mathbf{Q}_{n}\big( \mathbf{I} - (\mathbf{I} - h\gamma \mathbf{H}_n)^{-1}\big)\mathbf{Q}_{n}^T \mathbf{F}_i
            - \sum_{j=1}^{i-1} \frac{\gamma_{ij}}{\gamma} \mathbf{k}_j$ \label{a:l:inv}
    \EndFor
    \State $\mathbf{u}_* = \mathbf{u}_{n} + h_n \sum_{j=1}^{s} b_j \mathbf{k}_j$
    \State $\widehat{\mathbf{u}}_* = \mathbf{u}_{n} + h_n \sum_{j=1}^{s} \hat{b}_j \mathbf{k}_j$
    \State $err_* = |(\widehat{\mathbf{u}}_* - \mathbf{u}_*) / (Rtol \cdot \mathbf{u}_* + Atol)|$
    \State $h_* = h_n \cdot \min\big\{5.0,\, \max\{0.2,\, 0.8 \cdot err_{n}^{\beta} / err_*^{\alpha} \big\} \big\}$
    \If {$err_* \le 1$}	\Comment{step accepted}
        \State $\mathbf{u}_{n+1} = \mathbf{u}^*$
        \State $t \gets t + h_n$
        \State $h_{n+1} = h_*$
        \State $n \gets n + 1$
    \Else	\Comment{step rejected}
    	\State $h_n = h_*$
    \EndIf
\EndWhile
\EndProcedure
\end{algorithmic}
\end{algorithm}

\begin{figure}[ht]
\begin{center}
 \subfigure[\label{f:CH4_Air_Ref} Auto-ignition process]
 {\includegraphics[width = 0.49\textwidth]{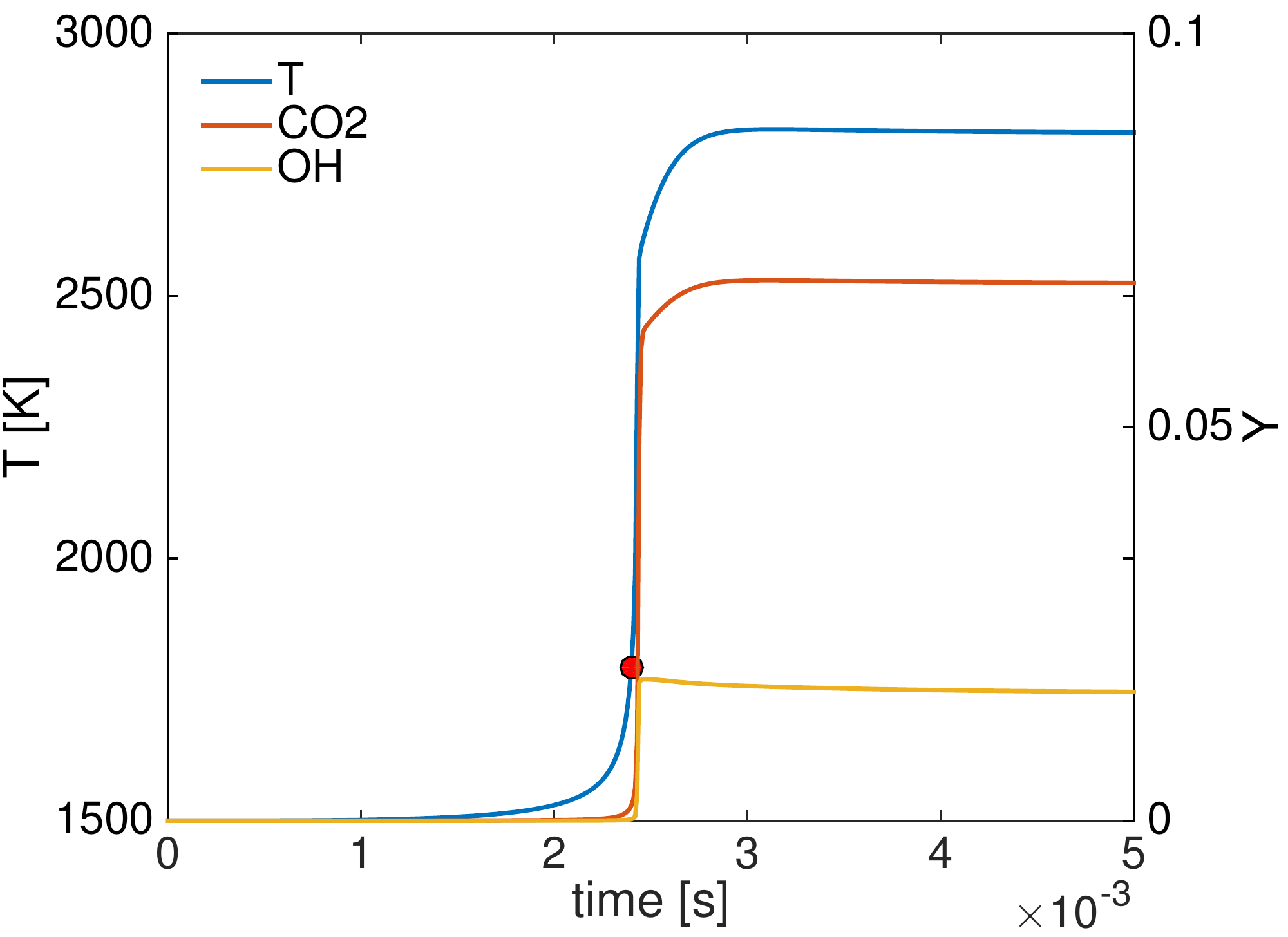}}
 \subfigure[\label{f:ROK4E_CH4_Air_Order} Order of convergence]
 {\includegraphics[width = 0.47\textwidth]{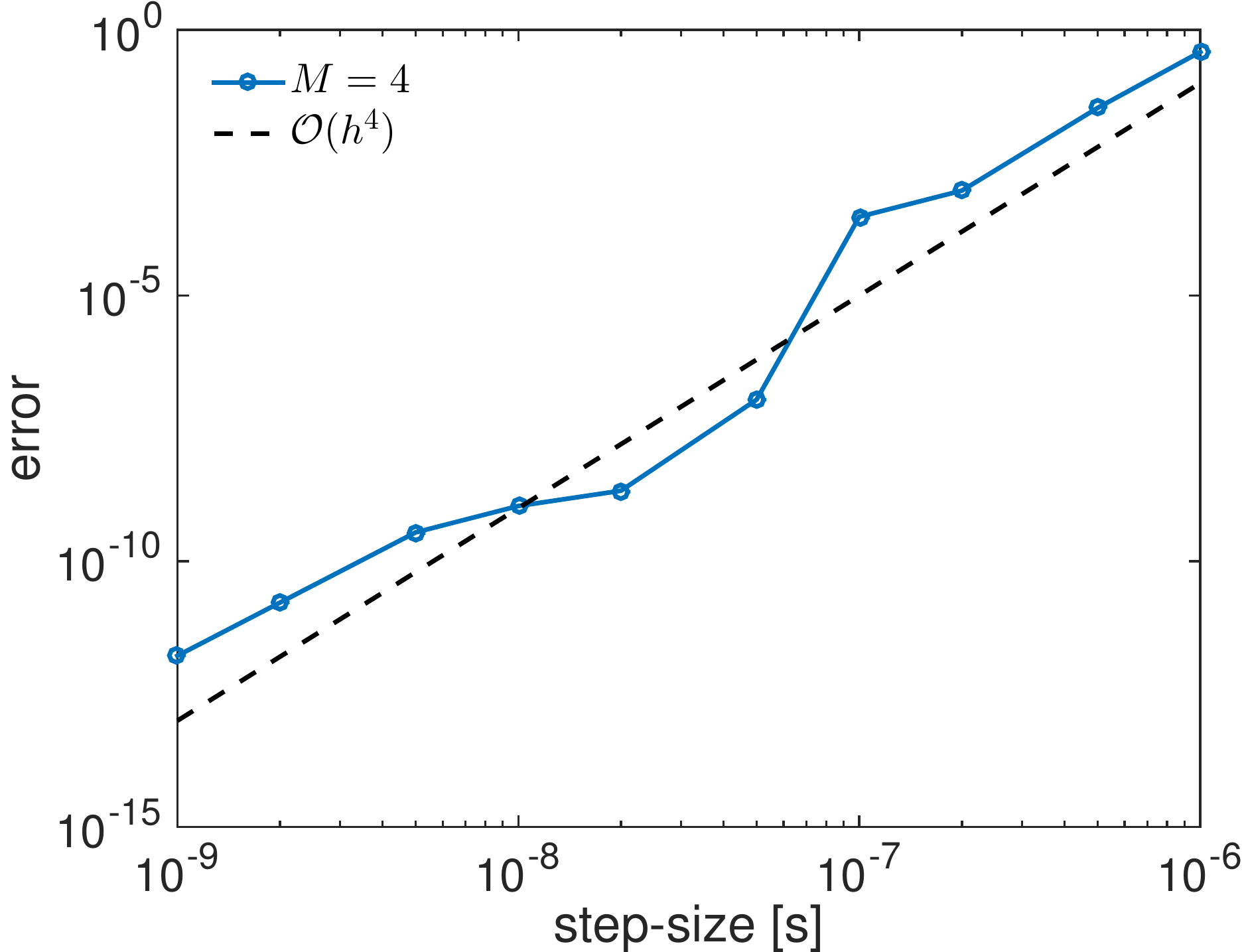}}
 \caption{\label{f:CH4_Air_SOLUT}
 Auto-ignition process for a stoichiometric mixture of \ce{CH_4}/air at $T = 1500 \, \si{\kelvin}$ and $P = 1 \, \text{atm}$ modeled by the GRI~$3.0$ mechanism. The reference convergence of order $4$ is plotted as the black dashed line.}
\end{center}
\end{figure}

\subsection{Numerical examples}
\subsubsection*{$0$-D methane/air reactor}
The first test case presented is a constant-volume $0$-D reactor of methane-air. This system, governed by the following equations
\begin{subequations}
\label{e:cv_reactor}
\begin{align}
	d_t Y_k &= \frac{1}{\rho}\dot{\omega}_{k} \\
    d_t T &= - \frac{1}{\rho c_v} \sum_{k}^{N_S} h_k \dot{\omega}_{k} ,
\end{align}
\end{subequations}
closely resembles the reaction part in the operator-splitting methods.

To begin with, the auto-ignition process of a \ce{CH_4}/air mixture modeled by the GRI~$3.0$ mechanism~\cite{GRI30} is considered. The initial temperature and pressure of the stoichiometric mixture are set to be $300 \, \si{\kelvin}$ and $1 \, \text{atm}$. 

The order of the ROK4E method is verified by measuring the relative error against the constant time-step size. The integrator is initialized with the reference solution obtained at $t = 2.4 \, \si{\milli\second}$, labeled by the red dot in Fig~\ref{f:CH4_Air_Ref}, and the integration is carried out over a period of $1 \, \si{\micro\second}$. The dimensionality of the Krylov-subspace set to be $M = 4$, corresponding to the minimal amount of implicitness while satisfying the order condition. As shown in Fig~\ref{f:ROK4E_CH4_Air_Order}, relative error reduces at the order of $4$, although slight degradation can be observed for very small time-step size due to the truncation error associated matrix-free approximation of the Jacobian-vector product.

The efficiency of the ROK4E method is assessed using the same configuration. In contrast to the convergence verification, adaptive step-size control is activated for this purpose. The integrations are carried out over intervals of size $h_{\text{CFL}}$ to mimic the usage scenario in a CFD solver. The results cover the range of $h_{\text{CFL}}$ between $10^{-9}$ to $10^{-5} \, \si{\s}$. For the ROK4E integrator, three levels of implicitness are tested with $M = 4, 6, 8$. Two additional integrators are also used for reference, one being the explicit Dormand–Prince (RKDP) method of order $5$ and the other being the implicit variable-order BDF method with maximum order of $5$. The \texttt{ROWPlus} implementation of the RKDP method and the \texttt{CVODE}~\cite{cohen1996cvode} implementation of the BDF method are chosen. The BDF method uses a numerical Jacobian paired with a dense linear solver for the modified Newton iteration. The evaluation of the chemical source term is performed by \texttt{Cantera}~\cite{goodwin_david_g_2017_170284}. The same level of tolerance, $Rtol = 10^{-4}$ and $Atol = 10^{-8}$, is specified for the ROK4E cases. The tolerance specifications for the BDF and RKDP integrators are adjusted such that the relative error is matched at $t = 2.4 \, \si{\milli\second}$.

\begin{figure}[ht]
\begin{center}           
 \subfigure[\label{f:CH4_Air_Cost} CPU time for $10^{-9} \le h_{\text{CFL}} \le 10^{-5} \, \si{\s}$]
 {\includegraphics[width = 0.49\textwidth]{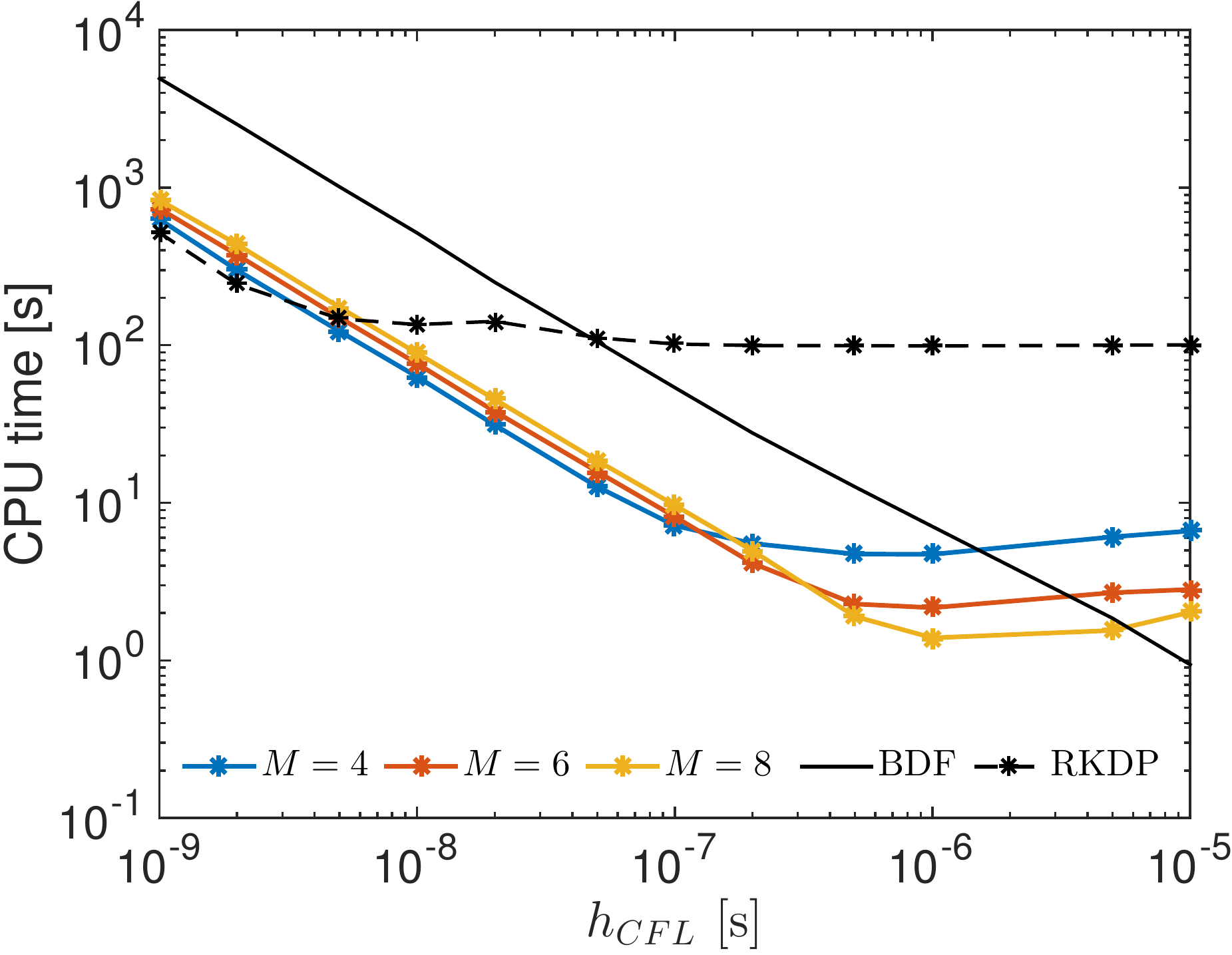}}
 \subfigure[\label{f:CH4_Air_Step_Size} Step-size at $h_{\text{CFL}} = 10^{-5} \, \si{\s}$]
 {\includegraphics[width = 0.48\textwidth]{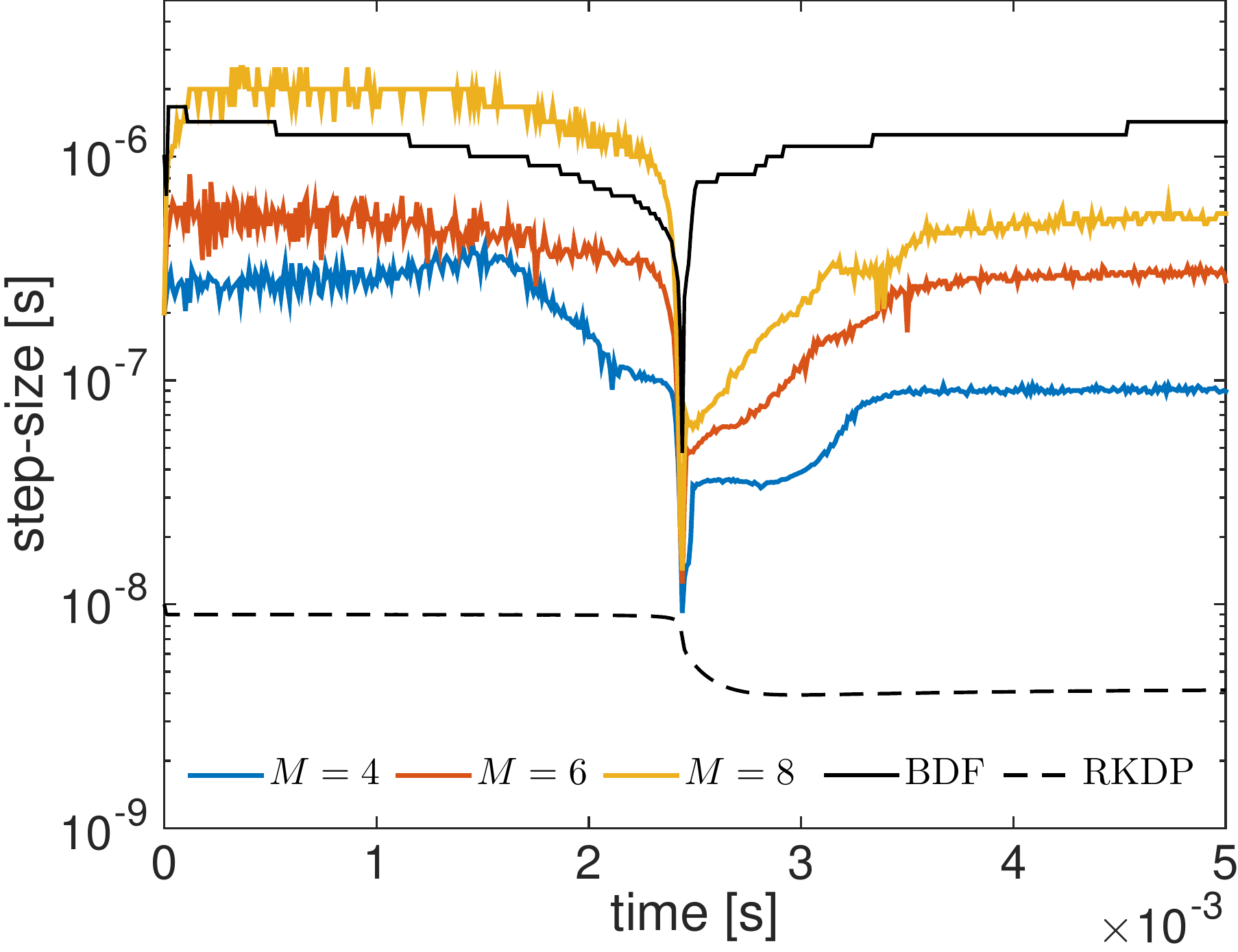}}
 \caption{\label{f:CH4_Air_Efficiency}
Comparison of the computational efficiency between ROK4E, BDF, and RKDP integrators using the auto-ignition calculation of a stoichiometric mixture of \ce{CH_4}/air at $T = 1500 \, \si{\kelvin}$ and $P = 1 \, \text{atm}$ modeled by the GRI~$3.0$ mechanism.}
\end{center}
\end{figure}

The the measured CPU time of different integrators as a function of $h_{\text{CFL}}$ is shown in Fig.~\ref{f:CH4_Air_Cost}. The semi-implicit ROK4E methods have a cost advantage over the fully implicit BDF method and the fully explicit RKDP method for $h_{\text{CFL}}$ ranging from $10^{-8}$ to $10^{-6} \, \si{\s}$, which covers the typical operating conditions for high-fidelity CFD solvers. To better understand the source of the benefit in efficiency, the step-sizes taken by each integrator for $h_{\text{CFL}} = 10^{-5}$ are shown in Fig~\ref{f:CH4_Air_Step_Size}. For the most part of auto-ignition process, the step-size is limited by the stability constraints. As a result, larger step-size can be taken when more implicitness is introduced to the integrator, which corresponds to the increasing value of $M$. The largest step-size is realized by the fully implicit BDF method, which also has the highest computational cost per-step. As the $h_{\text{CFL}}$ decreases, the advantage of having a larger step-size diminishes and methods with more implicitness becomes less efficient. It is worth noting that the ROK4E method with minimal implicitness is able to increase the step-size by over an order of magnitude, compared to the fully explicit RKDP method, while only requiring one more RHS evaluation per-step. 

\begin{figure}[h]
\begin{center}           
 \includegraphics[width = 0.50\textwidth]{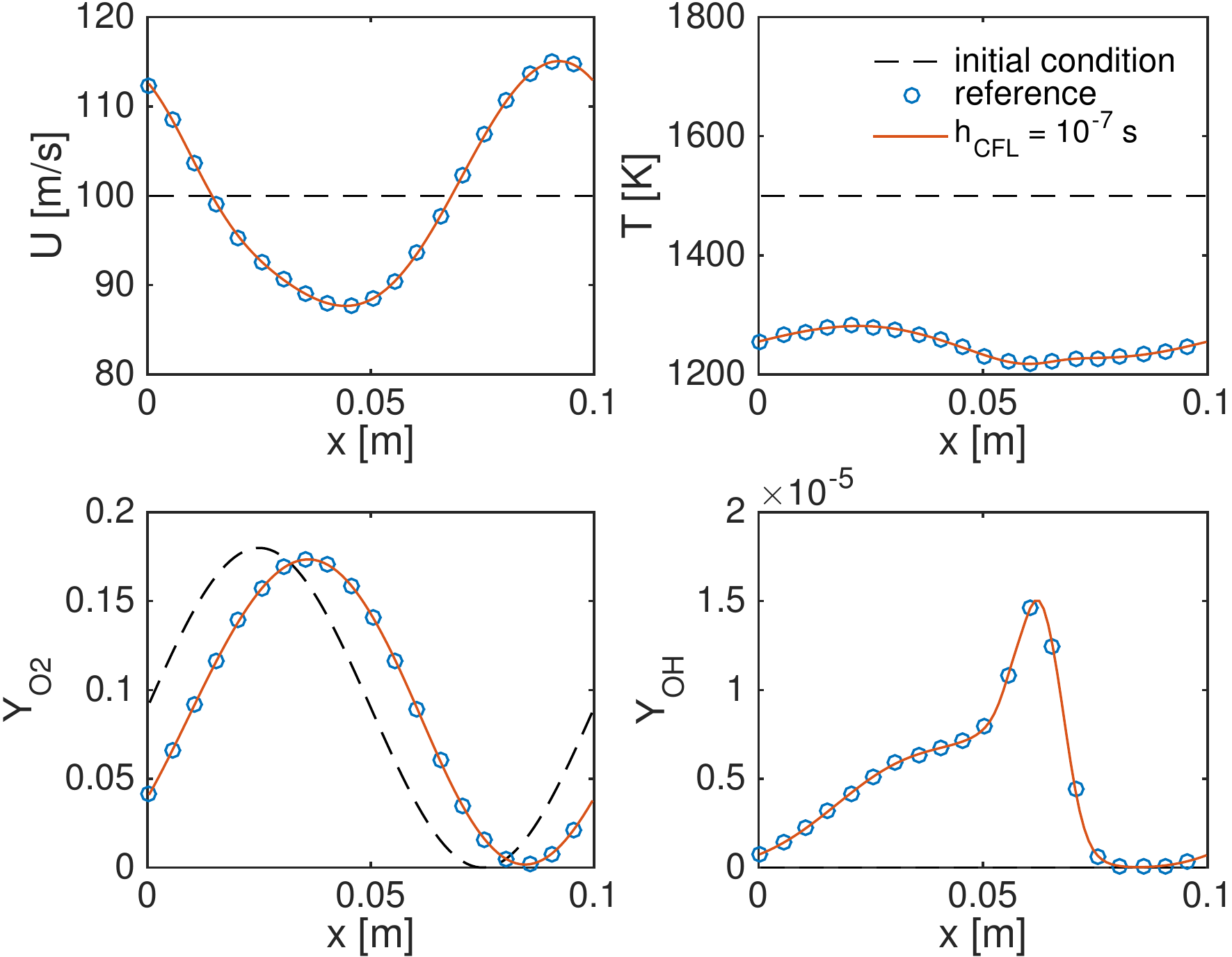}
 \caption{\label{f:OneD_DME_SOL}
Profiles of velocity, temperature, $\ce{O_2}$ and $\ce{OH}$ mass fractions at $t = 10^{-4} \si{\s}$. The initial condition is in black dashed line.}
\end{center}
\end{figure}

\begin{figure}[h]
\begin{center}           
 \includegraphics[width = 0.50\textwidth]{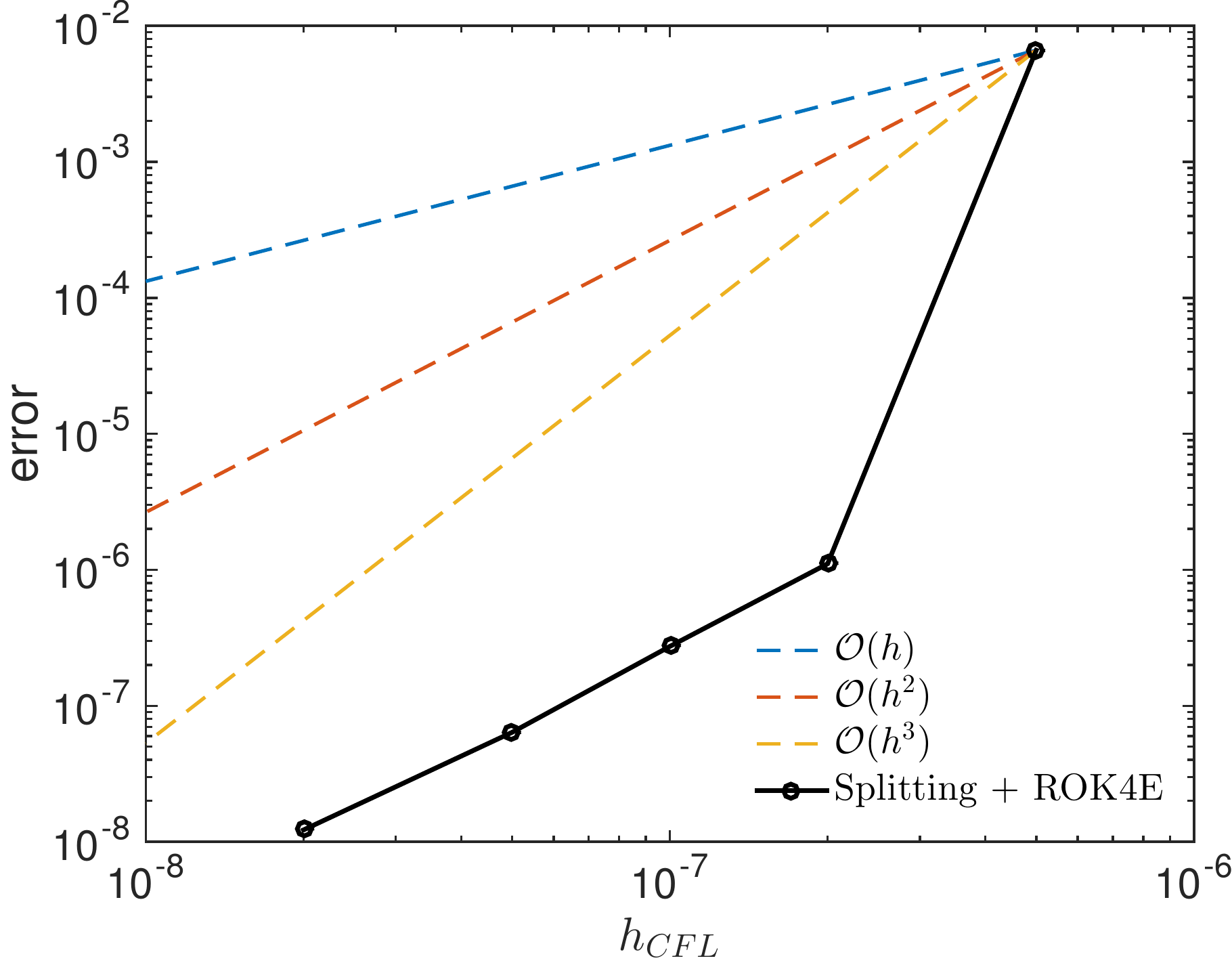}
 \caption{\label{f:OneD_DME_ERR}
Relative error of temperature as a function of CFD step-size $h_{\text{CFL}}$}.
\end{center}
\end{figure}

\subsubsection*{$1$-D DME/air flame}
The second test case considers a $1$-D DME/air flame. The full set of reactive Navier-Stokes equations is solved using the combination of the ROK4E integrator and the \textit{simpler} balanced splitting method. The combustion chemistry is modeled by a 30-species mechanism is used, which is reduced from a 39-species skeletal mechanism~\cite{LU_LAW_LQSSA_2006}. Since the thermal and chemical properties are explicit in temperature instead of internal energy, an auxiliary temperature equation is solved in addition to the total energy equation to avoid the iterative procedures of obtaining temperature from conservative quantities. As such the reaction system integrated by the ODE solver can be written as
\begin{equation}
\label{e:chemical_source_balanced}
	d_t	\begin{bmatrix}
            \mathbf{U} \\
            T \\
        \end{bmatrix} 
=	\begin{bmatrix}
            \mathbf{\mathcal{R}} \\
            0 \\
        \end{bmatrix} 
+ 	\begin{bmatrix}
            \mathbf{\mathcal{T}}_n \\
            0 \\
    \end{bmatrix} 
+ 	\begin{bmatrix}
			\mathbf{0} \\
            \dot{\omega}_T \\
    \end{bmatrix} 
\end{equation}
where $\mathbf{U}$ denotes the conservative variables as in Eq.~\ref{e:conserv_varaibles}, $\mathbf{\mathcal{R}}$ is the chemical source term as in Eq.~\ref{e:chemical_source}, $\mathbf{\mathcal{T}}_n$ is the balancing constant obtained from the transport term of Eq.~\ref{e:convection_diffussion}. The source of temperature, denoted by $\dot{\omega}_T$, is in the form of
\begin{equation}
\label{e:aux_t_equation}
\dot{\omega}_T = \frac{1}{\rho c_v} 
	\bigg(
    	-\sum_{k}^{N_S} h_k d_t\big(\rho Y_{k}\big)
    	+ \frac{1}{2} ( \mathbf{u} \cdot \mathbf{u} ) d_t \big(\rho \big) 
    	- \mathbf{u} \cdot d_t \big(\rho \mathbf{u} \big)
    	+ d_t \big( \rho E\big)
    \bigg) .
\end{equation}

The auxiliary temperature is initialized at the beginning of the reaction sub-step using the conservative variables and is used only within the sub-step for evaluating thermal and chemical properties. Hence, the conservation property of the system is not affected. Furthermore, the auxiliary temperature equals that obtained from the conservative variables if Eq.~\ref{e:aux_t_equation} is integrated exactly, which means that the potential error introduced by this treatment has the same order of convergence as the integration method.

The initial mixture consists of four species: $\ce{O_2}$, $\ce{N_2}$, $\ce{CH_3OCH_3}$ (DME), and $\ce{H}$ radical. The mass fraction of $\ce{O_2}$, $\ce{CH_3OCH_3}$, and $\ce{H}$ follows out-of-phase sine waves. The initial composition profile is convected with a uniform velocity $u = 100 \, \si{\m / \s}$ in a periodic domain. The pressure and temperature field is also uniformly at $P = 1 \, \text{atm}$ and $T = 1200 \, \si{\kelvin}$. The solution at $t = 10^{-4} \si{\s}$ is shown in Fig.~\ref{f:OneD_DME_SOL} in comparison to the initial profile. The solid red line is obtained with $h_{\text{CFL}} = 10^{-7} \si{\s}$ and the referenced solution in blue circle is calculated using non-split SSP-RK3 with $h_{\text{CFL}} = 10^{-9} \si{\s}$. In Fig.~\ref{f:OneD_DME_ERR}, the expected second order convergence in $T$ is demonstrated.

\subsubsection*{$3$-D DME/air flame}

\begin{figure}[h]
\begin{center}           
 \includegraphics[trim={8cm 8cm 8cm 8cm},clip,width=0.6\textwidth]{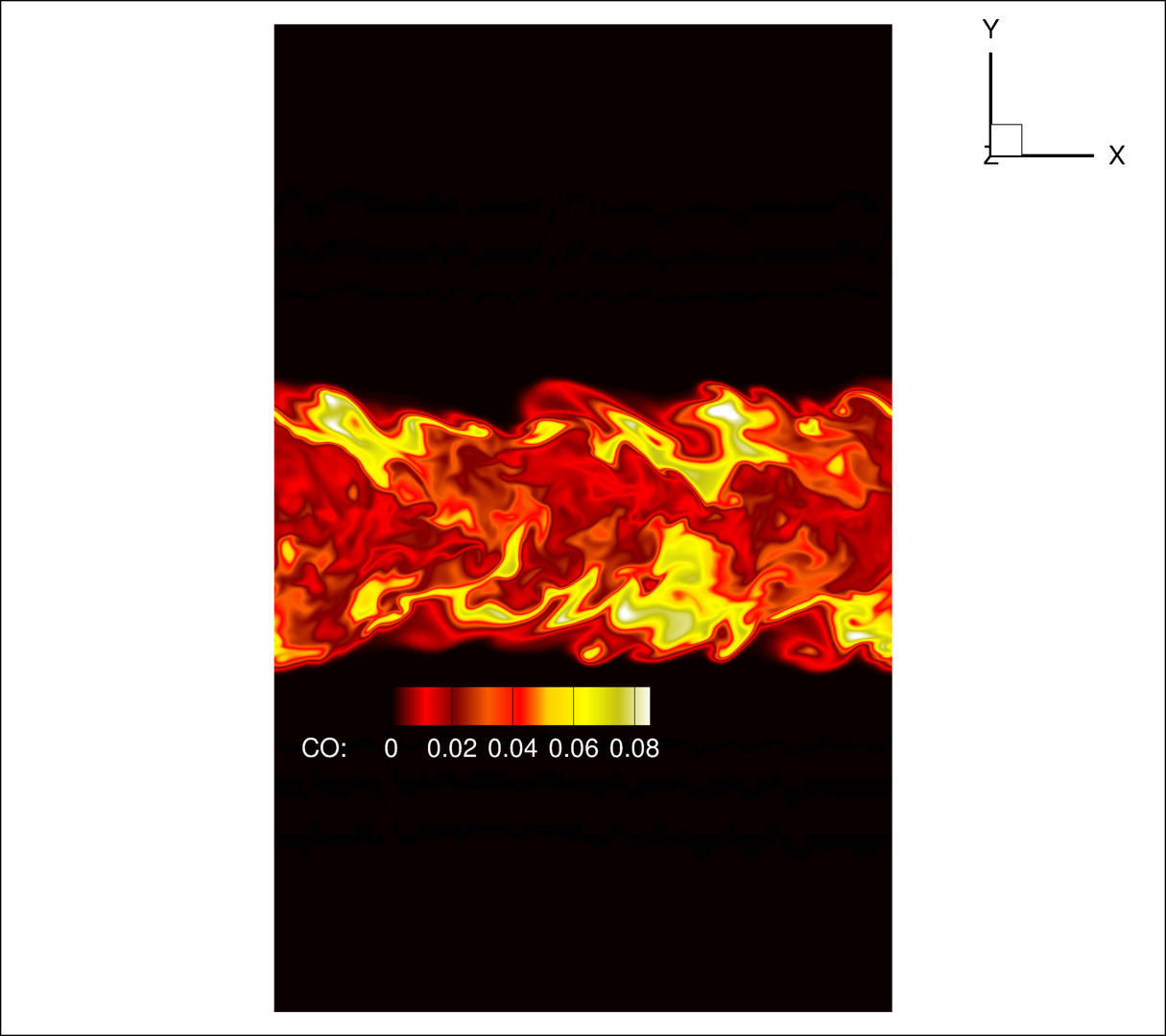}
 \caption{\label{f:3D_DME_CO}
 	Instantaneous field of $\ce{CO}$ mass fraction for a temporally evolving planar slot jet DME/air flame.}
\end{center}
\end{figure}

\begin{figure}[h]
\centering
\includegraphics[height=0.38\textwidth]{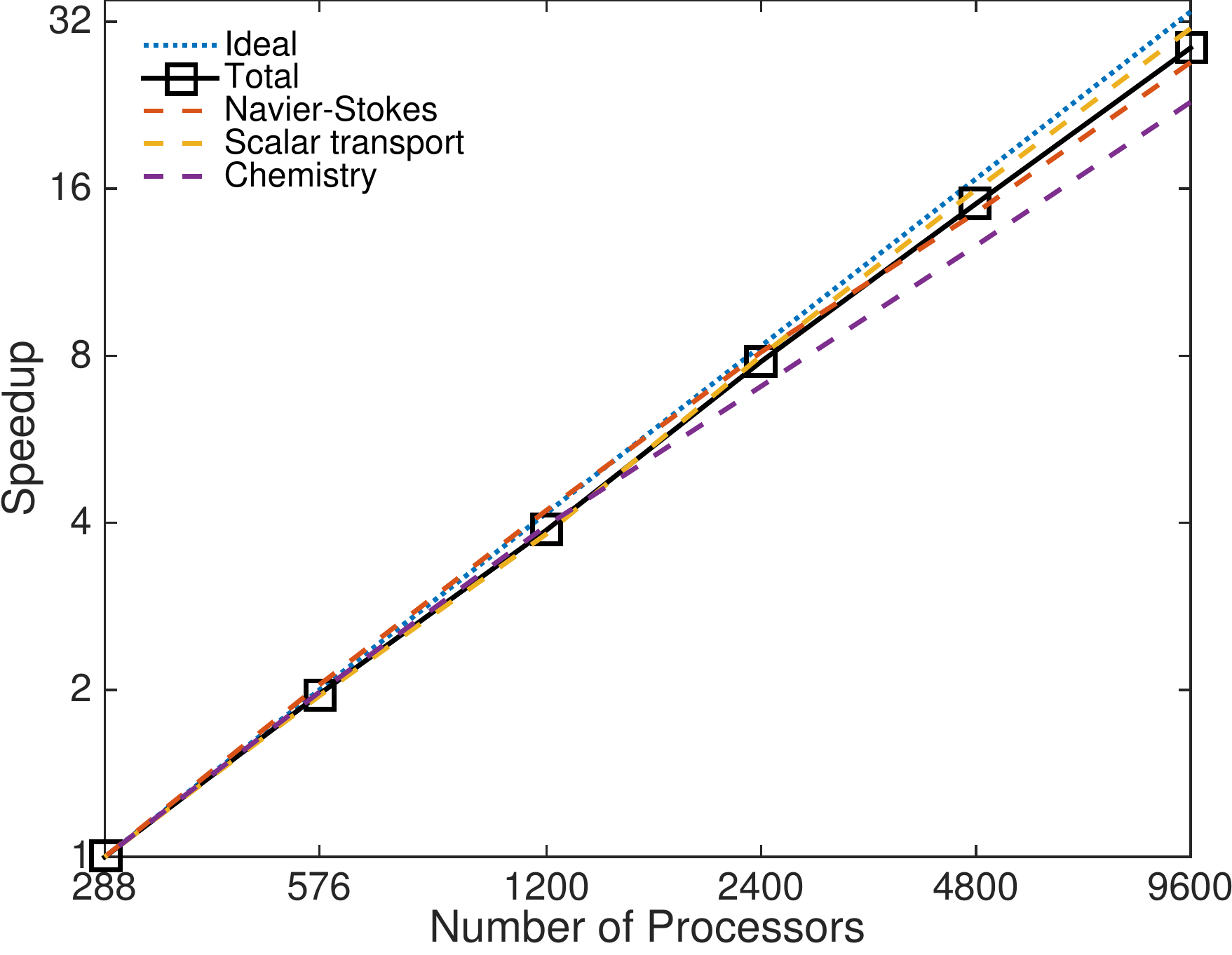}
\includegraphics[height=0.39\textwidth]{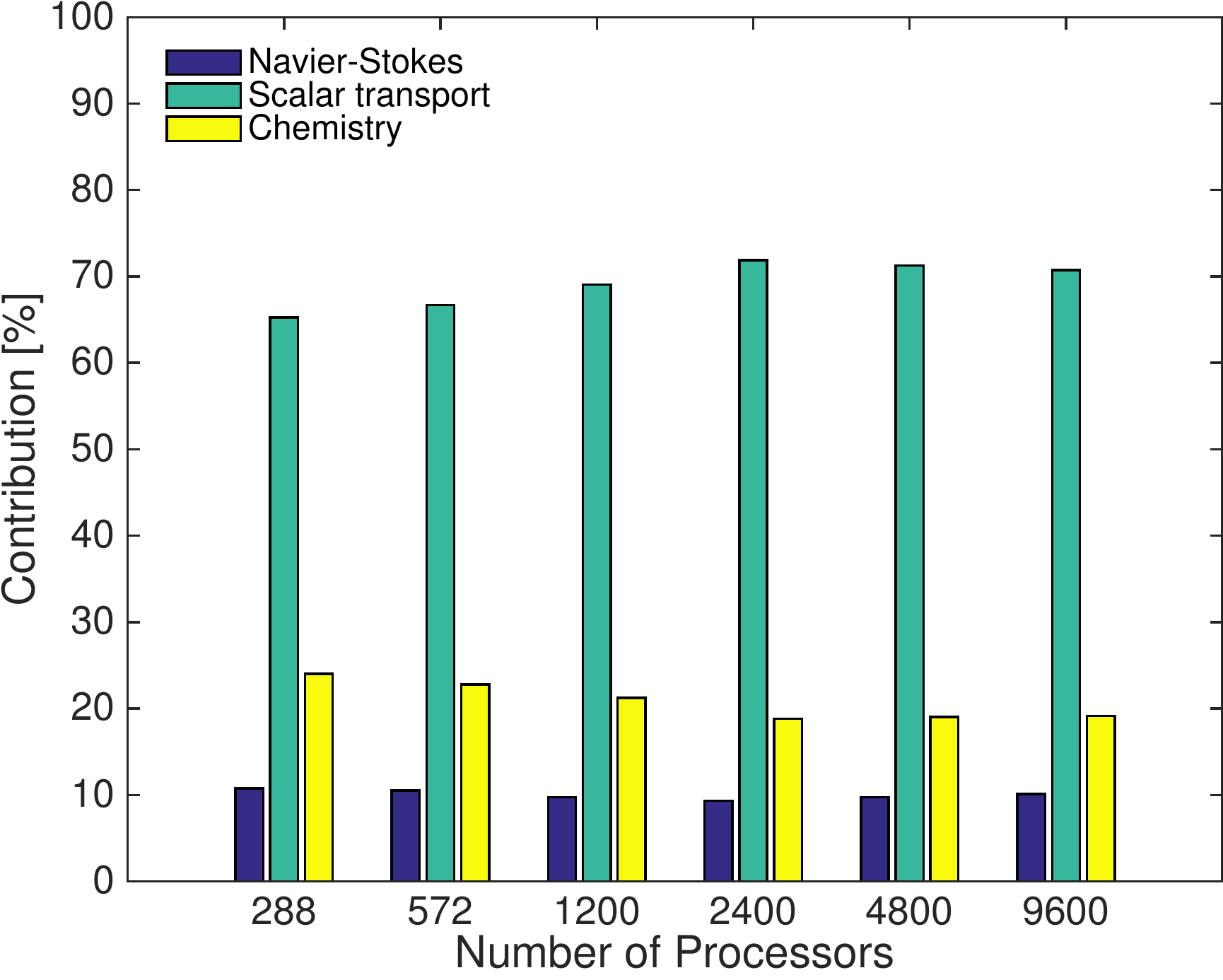}
\caption{\label{fig:3d_DME_SPEED}Strong scaling results of the CharLES$^x$ code on Haswell processors for DME/air flame with a 30-species reduced mechanism. Left: speed-up for the total code, the Navier-Stokes part, the scalar transport part, the combustion chemistry part, compared to the ideal scaling. Right: contribution of each part to the total CPU consumption.}
\label{speedup_plei}
\end{figure}

Finally, a full $3$-D calculation is performed for a turbulent DME/air jet flame. The configuration and the operating condition of the temporally evolving planar slot jet flame follow the DNS calculated by Bhagatwala et al.~\cite{bhagatwala2015numerical}, with the initial profile chosen from a near-extinction laminar strained flame solution. The resolution is reduced by a factor of $3$ in all directions compared with the DNS, leading to a mesh of 27 million control-volumes. The boundary conditions are periodic in the stream-wise and span-wise directions, and non-reflecting outflow in the transverse direction. An instantaneous field of $\ce{CO}$ mass fraction is show in Fig.~\ref{f:3D_DME_CO}. The $3$-D calculation is performed using an unstructured finite-volume solver, CharLES$^x$~\cite{KHALIGHI_NICHOLS_HAM_LELE_MOIN_AIAA2011,ma2017entropy,ma2017flamelet,wu2017mvp}, developed at CTR.

The computational efficiency of the proposed methods are evaluated by performing the simulation using up to 9600 Haswell processors. The strong scalability is tested by increasing the number of utilized CPUs for simulations of the same size. The results are shown in Fig.~\ref{fig:3d_DME_SPEED}, with all three major components of computational, namely the Navier-Stokes equations, scalar transport equations, and the combustion chemistry, achieving speed-up close to the ideal scaling. The relative contributions of these three parts are also calculated. The scalar transport accounts for 70-80\% of the total cost. The integration of the combustion chemistry accounts for as low as 20\% of the total cost, demonstrating the high efficiency of the time integration methods.

\section{Conclusion}
In this work, a combination of a steady-state preserving operator splitting method and a semi-implicit integration scheme is proposed for efficient time stepping of reactive turbulent simulations with stiff chemistry. The \textit{Simpler} balanced splitting method is constructed, with improved stability properties and reduced computational cost. The method is shown to be capable of stable and accurate prediction of ignition and extinction for reaction-diffusion systems near critical conditions. The ROK4E scheme is designed for semi-implicit integration of spatially independent reacting systems. Being a Rosenbrock-Krylov method, ROK4E utilizes the low-rank approximation of the Jacobian to reduce the cost for integrating the system of ODEs that have relative few stiff components. The efficiency of the scheme is further improved via the careful choice of coefficients to require three RHS evaluations over four stages. Combing these two methods, efficient calculation is achieved for large-scale parallel simulations of turbulent flames.    

\section*{Acknowledgments}
\label{Acknowledgments}
Financial support through NASA Award No. NNX15AV04A is gratefully acknowledged. Resources supporting this work were provided by the  High-End Computing (HEC) Program at NASA Ames Research Center.










\end{document}